\documentclass[twocolumn]{aastex63}

\usepackage{natbib}
\usepackage{color}

\usepackage{epstopdf}

\accepted{by ApJ 9th April 2020}

\shorttitle{Spectral signatures of chromospheric condensation}
\shortauthors{Graham et al.}

\begin{document}


\title{Spectral signatures of chromospheric condensation in a major solar flare}


\correspondingauthor{G. Cauzzi}
\email{gcauzzi@nso.edu}

\author{David R. Graham}
\affiliation{Bay Area Environmental Research Institute, Moffett Field, CA 94035, USA}

\author{Gianna Cauzzi}
\affiliation{National Solar Observatory, University of Colorado Boulder, 3665 Discovery Drive, Boulder, CO 80303, USA}
\affiliation{INAF-Osservatorio Astrofisico di Arcetri, I-50125 Firenze, Italy}

\author{Luca Zangrilli}
\affiliation{INAF-Osservatorio Astrofisico di Torino, I-10025 Pino Torinese, Italy}

\author{Adam Kowalski}
\affiliation{National Solar Observatory, University of Colorado Boulder, 3665 Discovery Drive, Boulder, CO 80303, USA}
\affiliation{Department of Astrophysical and Planetary Sciences, University of Colorado, Boulder, 2000 Colorado Ave, CO 80305, USA}
\affiliation{Laboratory for Atmospheric and Space Physics, University of Colorado Boulder, 3665 Discovery Drive, Boulder, CO 80303, USA}

\author{Paulo Sim\~oes}
\affiliation{Centro de R\'adio Astronomia e Astrof\'isica Mackenzie, Escola de Engenharia, Universidade Presbiteriana Mackenzie, Brazil}
\affiliation{SUPA School of Physics and Astronomy, G12 8QQ, University of Glasgow, UK}
\and
\author{Joel Allred}
\affiliation{NASA/Goddard Space Flight Center, Code 671, Greenbelt, MD 20771, USA}




\begin{abstract}
We study the evolution of chromospheric line and continuum emission during the impulsive phase of the X-class SOL2014-09-10T17:45 solar flare. We extend previous analyses of this flare to multiple chromospheric lines of Fe {\sc i}, Fe {\sc ii}, Mg {\sc ii}, C {\sc i}, and Si {\sc ii} observed with IRIS, combined with radiative-hydrodynamical (RHD) modeling. For multiple flaring kernels, the lines all show a rapidly evolving double-component structure: an enhanced, emission component at rest, and a broad, highly red-shifted component of comparable intensity. The red-shifted components migrate from 25-50  km s$^{-1}$ towards the rest wavelength within $\sim$30 seconds.

Using Fermi hard X-ray observations, we derive the parameters of an accelerated electron beam impacting the dense chromosphere, using them to drive a RHD simulation with the RADYN code. As in \cite{2017ApJ...836...12K}, our simulations show that the most energetic electrons penetrate into the deep chromosphere, heating it to T$\sim$10,000 K, while the bulk of the electrons dissipate their energy higher, driving an explosive evaporation, and its counterpart condensation --- a very dense (n$_e \sim 2 \times 10^{14}$ cm$^{-3}$), thin layer (30--40 km thickness), heated to 8--12,000 K, moving towards the stationary chromosphere at up to 50 km s$^{-1}$.

The synthetic Fe {\sc ii} 2814.45\AA\ profiles closely resemble the observational data, including a continuum enhancement, and both a stationary and a highly red-shifted component, rapidly moving towards the rest wavelength. Importantly, the absolute continuum intensity, ratio of component intensities, relative time of appearance, and red-shift amplitude, are sensitive to the model input parameters, showing great potential as diagnostics.

\end{abstract}

\keywords{Sun: activity - Sun: chromosphere - Sun: flares - Sun: transition region - Sun: UV radiation - Sun: X-rays, gamma rays}

\section{Introduction} \label{sec:intro}

Solar flares bear on multiple aspects of plasma physics, including the physics of magnetic reconnection, the transport of energy from the corona to the lower atmosphere, the production of coronal mass ejections driving space weather, etc. \citep{2011LRSP....8....6S,2017LRSP...14....2B}. The solar flare paradigm remains widely accepted as a template for magnetic activity on other stars \citep[e.g.][]{2017ApJ...851...91N}, thus a full understanding of the phenomenon has a large relevance for astrophysics in general.

In particular, the chromosphere remains a crucial element to understand the development of flares. While the primary
energy release mechanism is governed by magnetic reconnection in the corona, the chromosphere is the location where this
energy is ultimately deposited, either via conduction or accelerated particles; this gives rise to the largest contribution to the flare radiative output \citep{2011SSRv..159...19F,2014ApJ...793...70M}, and, via chromospheric evaporation, to the necessary mass and energy to fill the flaring loops with plasma at temperatures in excess of 10 MK \citep{1985ApJ...289..414F}.  

After a long period of relative neglect, the past decade has seen a renewed interest into the chromospheric
response to flares. This has been driven partly by the availability of novel instrumentation, such as the imaging spectro-polarimeters IBIS \citep{2006SoPh..236..415C} and CRISP \citep{2008ApJ...689L..69S}, as well as the recent  {\it Interface Region Imaging Spectrograph} \citep[IRIS,][]{2014SoPh..289.2733D}, which has been utilized in numerous flare studies. On the other hand, building on the pioneering  work  
of Fisher and colleagues in the 1980's \citep{1985ApJ...289..414F, Fisher:1986aa, Fisher:1987aa, 1989ApJ...346.1019F}, several groups have developed radiative hydrodynamical (RHD) simulations that aim at understanding 
the effects of a strong and sudden heating burst affecting the lower solar atmosphere
\citep{2005ApJ...630..573A, 2015ApJ...809..104A, 2015ApJ...808..177R, 2015ApJ...813..133R,2016IAUS..320..233H}. %
The complexities of fully accounting for hydrodynamical (including the presence of shocks) and radiative transfer effects in 
determining the chromosphere's response to flaring limit the simulations to the 1-D case only;
this however is usually well justified by the control that magnetic fields exert to plasma dynamics \citep[cf., e.g.][]{2015ApJ...809..104A,2015ApJ...813..133R}. 

The flaring chromospheres resulting from the models can then be compared with proper observational diagnostics to gauge
their realism; most often, this is accomplished by using radiative transport codes such as RH \citep{2001ApJ...557..389U} to synthesize relevant 
chromospheric lines (including the resonance lines of Mg~{\sc ii} h\&k, Ca~{\sc ii} H\&K, or H$\alpha$) in snapshots of 
the simulations' output, to provide a direct comparison with observables. Some of the relevant mechanisms influencing the formation 
of these strong, optically thick chromospheric spectral lines in flaring conditions remain an active subject of study,  including e.g. the effects of partial frequency redistribution for the Mg~{\sc ii} resonance lines \citep{2019ApJ...883...57K}, or the role of enhanced densities on the
broadening of the Hydrogen Balmer lines \citep{2017ApJ...837..125K}. Still, this method offers the best diagnostic possibilities, especially when the temporal evolution is factored in the comparison, as the dynamics of the flaring plasma are extremely
sensitive to the details of heating and energy transport \citep{2015ApJ...808..177R, 2015A&A...582A..50K, 2016ApJ...827..145R,2017ApJ...836...12K}.

Comparisons of this kind have been attempted many times in the past using strong optical lines \citep[e.g.][]{1990ApJ...348..333C,1992A&A...256..255F,2002A&A...387..678F}, and continued to date with data at much higher spatial resolution and spectral coverage \citep{2015ApJ...798..107K,2015ApJ...813..125K,
2016ApJ...827...38R,2017ApJ...847...48H,2017ApJ...836...12K, 2019ApJ...878..135K}. Still,
fully resolved -- spatially, temporally, spectrally -- observations of the impulsive phase of flares remain scarce, owing to the difficulty of positioning the slit of a spectrometer exactly at the flare footpoints, and exactly at the time of energy deposition in the lower atmosphere. As numerical simulations mostly concentrate on the very first instants ($\le 100$ s) of the flare's development, comparisons with observations have often been less than optimal. 

Recent observations of flare dynamics, obtained at high spatial and temporal resolution with the IRIS spacecraft and other
instruments, are much improving this situation \citep{2018ApJ...867...85B, 2018ApJ...856...34T, 2018SciA....4.2794J, 2019ApJ...879L..17P}. One of the best examples to date of chromospheric dynamics and condensation during a flare has been presented in \citet[][hereafter Paper I]{Graham:2015aa}, who exploited a unique sit \& stare, high cadence (9.4 s) IRIS dataset obtained during a strong flare, to reveal clear signatures of chromospheric condensation in multiple footpoints sources. The fortuitous development of the flare along the spectrograph's slit allowed the authors to uncover
the quantitative similarity of the dynamical behavior of a number of flaring pixels, hinting at the possibility of spatially resolved {\em elementary} flare kernels over sub-arcsec spatial scales. The use of imaging spectrometers working in the optical range appears also very promising: \cite{2019A&A...621A..35L} recently presented results obtained with CRISP, using the He~{\sc i}~${\rm D}_3$ line to study the chromospheric
dynamics in a small C-class flare. The large field of view afforded by the instrument made possible the spatial characterization of condensation motions, that were found in the leading edge of ribbons, as previously reported by \citet[][ albeit at a much lower spatial and temporal resolution]{1997A&A...328..371F}. Further, the condensation is found to decay from its maximum value of $\sim$ 60 km s$^{-1}$ down to zero in around 60 seconds, a very similar evolution to what reported in Paper I using the Mg~{\sc ii} 2791.6\AA~subordinate line.

An interesting result of \citet{2019A&A...621A..35L} is that the interpretation of their spectro-polarimetric He~{\sc i} data
in the flaring footpoints requires the use of two separate chromospheric components (``slabs''), one related to the condensation, and another one
pertaining to a deeper layer, where the spectral component appears enhanced with respect to quiet conditions, and modestly blue-shifted. This second component starts being visible about one time step (15 s) 
after the condensation's maximum red-shift, and has been interpreted by the authors as due to shock heating (and possible 
rebound) of the undisturbed chromosphere by the moving condensation. A similar scenario has been presented
by \citet{2017ApJ...836...12K}, using IRIS observation of Fe~{\sc ii} lines in flaring kernels: the authors showed
how in some instances these chromospheric spectral lines were clearly 
composed of two distinct spectral components, one at the rest wavelength and one strongly red-shifted. However, the
authors offer a fairly different interpretation than \citet{2019A&A...621A..35L}: using 
RADYN flare simulations \citep{2015ApJ...809..104A}, constrained by the hard X-ray (HXR) observations obtained with {\it RHESSI}, they could reproduce the observed shapes of the Fe~{\sc ii} lines as due to concomitant effects of the energy delivered by a beam of accelerated
electrons. In particular, while the low energy electrons (E $\sim$ 25 -- 50 keV) in the beam are responsible for
creating the evaporation and heating the chromospheric condensation -- producing a strongly red-shifted line, the 
higher energy electrons (E $\ge$ 50 keV) can penetrate in deeper, denser layers of the chromosphere and heat them, giving rise to the enhanced  stationary component. Key to understanding the flaring mechanism itself, is the fact that the relative timing, and strength, of these separate effects depends on the details of the heating input, including the hardness of the beam, and the duration of the input.

The IRIS flare studied by \citet{2017ApJ...836...12K} was observed in raster mode with a moderate cadence of 45~s, so that a full study of the temporal evolution of the chromospheric condensation could not be carried out. 
In the present paper we expand on these findings, using the unique dataset of Paper I to study the full 
temporal evolution of multiple chromospheric lines, in multiple flaring kernels, during the impulsive phase of the flare, with excellent spectral and temporal resolution. Following \citet{2017ApJ...836...12K},  we take advantage of co-temporary
 HXR data from Fermi \citep{2009ApJ...702..791M}, as well as  the IRIS Slitjaw Images (SJIs) to properly estimate the full parameters of the energy input. We further synthesize the Fe~{\sc ii} 2814\AA~ line for comparison with the data; 
as shown by \citet{2014ApJ...794L..23H, 2017ApJ...836...12K},  this line is an important diagnostic in flares because the intensity originates from a similar temperature range (with a broad peak around $T\sim12,000$ K) as hydrogen Balmer bound-free radiation that dominates the IRIS NUV range.

\section{The  SOL2014-09-10T17:45, X1.6 event}\label{sec:flare_description}

\begin{figure}
  \centering
  \includegraphics[width=10cm, trim = 0 60 0 0, , clip]{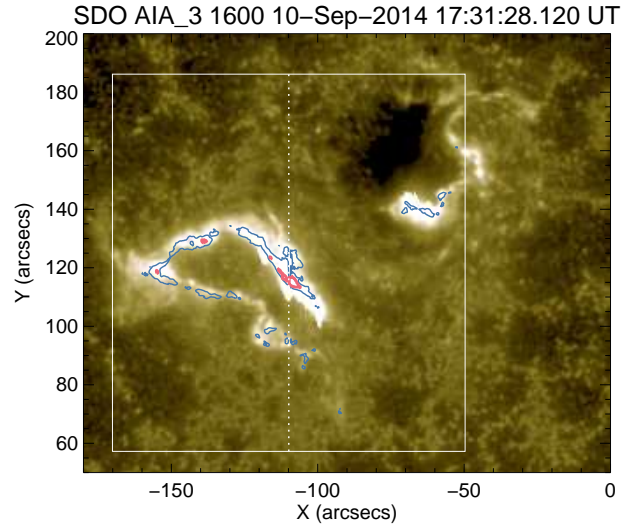}

   \caption{AIA 1600\AA\ image at maximum flare brightness. Contours in blue show the extent of the ribbon in IRIS 2796 \AA\ at 15\% of the maximum SJI intensity, while the brightest kernel is shown in pink, at the 50\% intensity level. The thin white box 
   indicates the IRIS SJI FOV, and the dotted line marks the IRIS slit position.}
     \label{fig:irisaia}
\end{figure}

The GOES X1.6 class flare SOL2014-09-10T17:45 developed in active region NOAA 12158 near disk center (N15E02). The complex two-ribbon structure encompassed both the western portion of the main, leading polarity sunspot, and a group of several smaller spots of following polarity embedded in the plage, as seen in Figure \ref{fig:irisaia}. The flare developed in a bursty manner, with UV kernels of various intensity appearing in rapid succession, in particular running north-east to south-west along the length of the plage ribbon. At the same time, the ribbons expanded roughly perpendicularly to their length, as in the canonical flare model.

Several authors have by now described many aspects of the flare, including the precursor phase \citep{2016ApJ...823L..19Z}, 
the eruption of a filament and slipping motion of flare loops \citep{2016ApJ...823...41D}, the presence of {\it quasi-periodic pulsations} \citep{2015ApJ...807...72L, 2015SoPh..290.3625S, 2017SoPh..292...11N}, and the occurrence of chromospheric evaporation \citep{2015ApJ...811..139T, 2015ApJ...813...59L} 
and its relation with chromospheric condensation (Paper I). Representative movies depicting the flare evolution are available in, e.g., \citet{2015ApJ...811..139T} and \citet{2016ApJ...823...41D}.

As discussed in the introduction, the present paper focuses on the chromospheric emission during the flare's impulsive phase, and its diagnostic potential with respect to radiative hydro-dynamical modeling. To this end, we will discuss mostly the hard X-ray (HXR) observations obtained by Fermi/GBM \citep{2009ApJ...702..791M} and the UV spectra and images acquired by IRIS \citep{2014SoPh..289.2733D}, in particular around the time of maximum IRIS UV emission. GOES soft X-ray (SXR) light curves and SDO/AIA images at various wavelengths will provide the necessary context.

\begin{figure}[!hbt]
  \begin{center}
   \includegraphics[width=8cm]{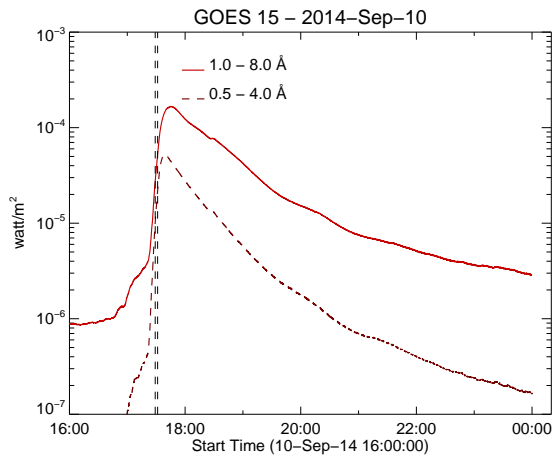}
    \caption{GOES 1--8  \AA\ flux over several hours around the event. The data utilized in this paper pertains to the very rapid
    impulsive phase, as framed by the dashed lines.}
    \label{fig:goes_intensity}
  \end{center}
\end{figure}

\begin{figure}[!hbt]
  \begin{center}
    \includegraphics[width=8cm]{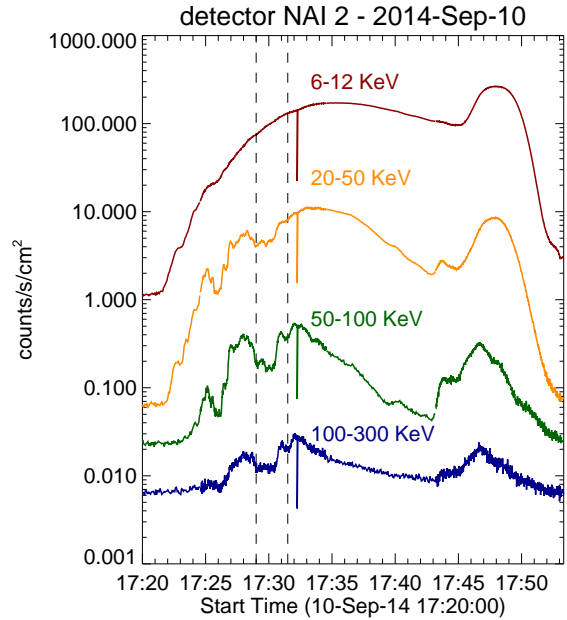}
    \caption{HXR detector count flux of the SOL2014-09-10T17:45 impulsive phase in several of the Fermi GBM energy bands. The 
    dashed lines frame the interval corresponding to the period of maximum UV emission analyzed in the paper.}
    \label{fig:hxr_intensity}
  \end{center}
\end{figure}

Figure \ref{fig:goes_intensity} shows the GOES SXR 1--8 \AA\ flux over most of the event's evolution. A modest increase in the SXR is visible already from about 16:50 UT, but the main impulsive phase starts only after 17:20 UT. The event then develops very rapidly, with the SXR flux peaking already at 
17:45 UT. The Fermi GBM count rates profiles, integrated in the energy bands 6-12, 20-50, 50-100, 100-300~keV, are shown in Figure~\ref{fig:hxr_intensity}. While RHESSI was in night-time during most of the flare, Fermi completely covered the impulsive phase, revealing a strong and rapid increase in the 20 -- 300 keV flux, starting from $\sim$ 17:22 UT and reaching its maximum around 17:36~UT. A second episode of enhanced HXR flux is also observed, between 17:43~UT and 17:50~UT. Further analysis on Fermi data is provided in Sect. \ref{sec:fermi}, in particular for the short time interval framed by the dashed lines in  Figure~\ref{fig:hxr_intensity}.

A unique set of observations was acquired by IRIS, that was running a flare watch program on NOAA 12158, using the sit and stare 
(SNS) mode and the standard flare line list (OBSID 3860259453).  High-cadence (9.4 s) flare spectra were obtained for many hours  
before the flare, and all the way to the end of the impulsive phase.  Several spectral windows were acquired within the 1332 - 1358 
\AA, 1389 - 1407 \AA\ (FUV), and 2783 - 2834 \AA\ (NUV) intervals \citep[see e.g. Fig. 3 in][]{2017ApJ...836...12K}. Exposure times 
were 8 s for both NUV and FUV channels up until the start of the flare; afterward, NUV exposures were reduced to 2.4 s to avoid 
saturation.  Simultaneous slit-jaw images (SJI) were obtained at a cadence of $\sim$19 s, alternating between two channels:  2796 \AA\ centered on Mg {\sc ii}; and 1400 \AA\ centered on Si {\sc iv}. The $120" \times 120"$  field of view (FOV) of IRIS is shown as a white box in Figure \ref{fig:irisaia}; the IRIS spectrograph slit is indicated with a dotted white line.

The peculiarities of the IRIS observations are many-fold: first,
the spectrograph's slit was positioned exactly over the plage flare ribbon, and intersected, among others, the strongest kernel within the whole flare (as observed in the SJIs). Second, the ribbon developed rapidly {\em along} the slit, providing multiple, 
independent instances of elementary flaring kernels (see Paper I for a more thourogh discussion). Third,  
the observations started well before the initiation of the flare, thus capturing the evolution of any given flaring area from their earliest moments. This is a very rare occurrence. Fourth, the 9.4 s cadence is among the highest ever achieved for UV flare spectra, and allows novel analyses of  the rapidly evolving chromospheric condensation. Finally, the strongest flaring kernels were so intense that many weak chromospheric lines clearly turned into emission, providing a range of complementary diagnostics beside the most-often used Mg~{\sc ii} and C~{\sc ii} lines. 

Figure \ref{fig:irisaia} shows the flare's ribbons at the time of maximum chromospheric emission, as determined from the 2796 \AA\ SJIs. The most intense brightening in the IRIS images is highlighted by a pink contour on the simultaneous AIA 1600 \AA\ 
image; at this time the kernel was at maximum brightness and directly sampled by the slit (the kernel can also be identified by the saturated region in Figure 1 of Paper I). The kernel was particularly bright between 17:29:48 - 17:31:41~UT; during this time its intensity 
accounted for 15-20\% of the total 2796 \AA\ SJI counts while accounting for only $\sim$1\% of the total SJI area. In the following we thus focus on this particular feature, by studying its chromospheric spectra and their evolution, as related to the HXR signatures. 
Given the high cadence of the HXR data, and the excellent sampling of IRIS, such a dataset is most appropriate to compare with results from hydrodynamical numerical simulations.

\section{Chromospheric dynamics (condensation)}\label{sec:chrom}

Aside from the most used Mg~{\sc ii} h \& k doublet, a multitude of chromospheric lines is found within the IRIS spectral range.  
Together, they can improve the interpretation of line shapes and shifts in terms of physical parameters of the flaring lower atmosphere.

In Paper I we showed the temporal evolution of chromospheric condensation in elementary flare kernels by analyzing the Mg~{\sc ii} 2791.6\AA\ subordinate line. In particular, we used the position of the line bisector at 30\% of the peak intensity level to estimate the amplitude of downflows.
Similarly to earlier observations and modelling \citep[e.g.][]{1989ApJ...346.1019F}, condensation velocities of up to $30-40~{\rm 
km~s^{-1}}$ were found, rapidly decaying to zero within $30-60$ s. While the evolution of the condensation velocity was extremely clear (see Figure 5, Paper I) we did not delve further into details of the line profiles themselves. It was, however, apparent that at several positions and times the Mg~{\sc ii} subordinate line profiles were strongly asymmetric, or even had distinct red-shifted 
components. Such a property was also noted by \citet{2015ApJ...811..139T} for several Si~{\sc ii} lines adjacent to the 
coronal Fe~{\sc xxi} line, but not analyzed further.

\begin{figure}
  \centering
 \includegraphics[width=9cm]{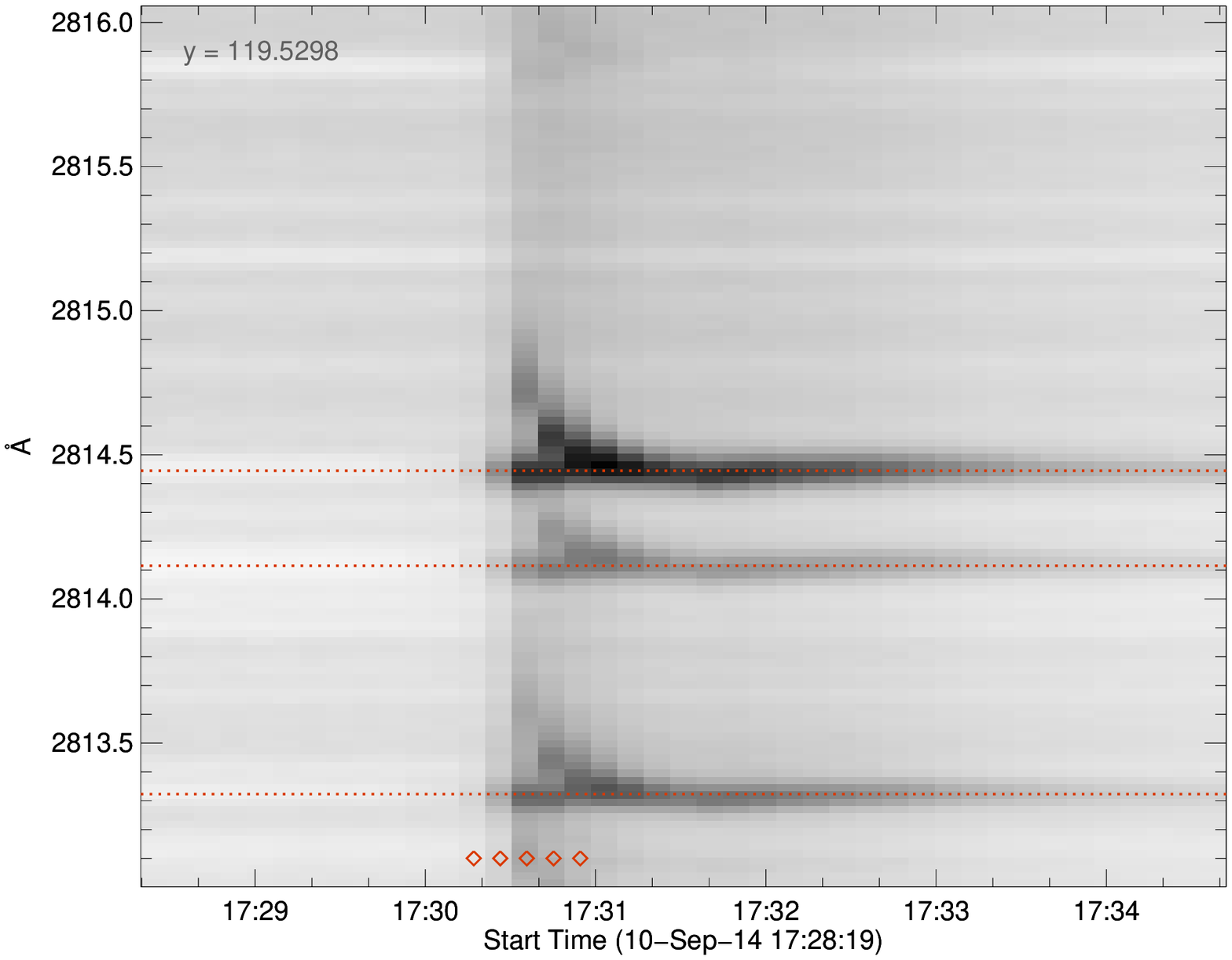}
  \includegraphics[width=9cm]{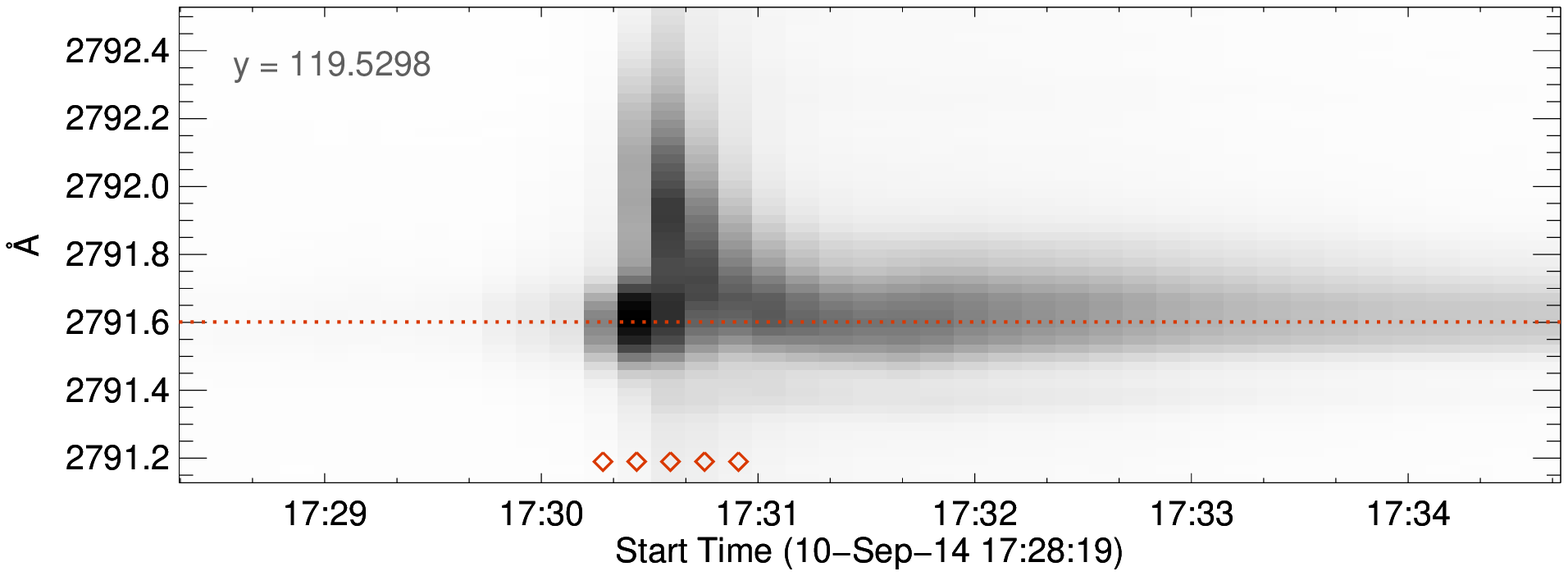}
    \includegraphics[width=9cm, trim = 0 0 0 0, , clip]{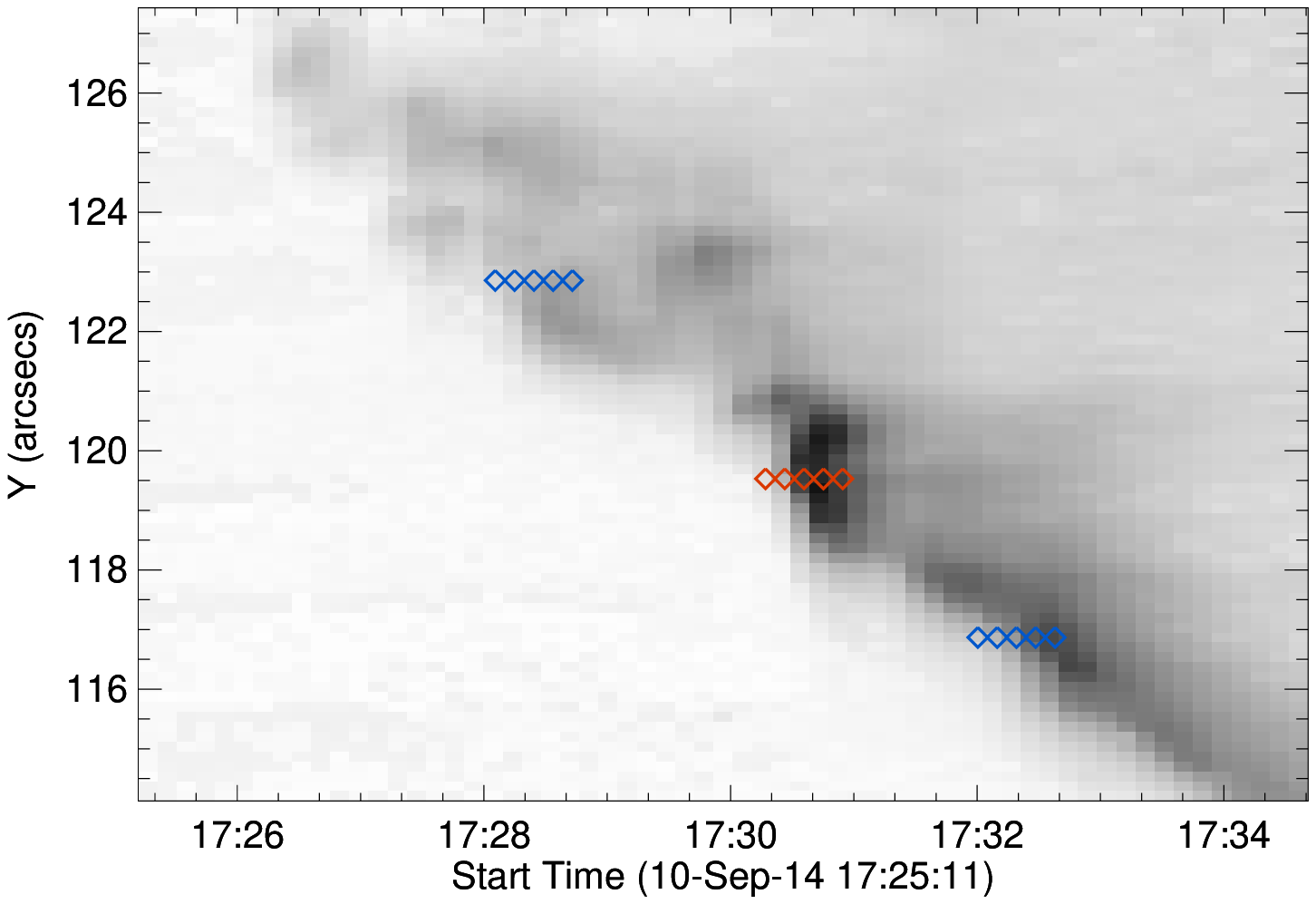}

   \caption{Temporal evolution of Fe~{\sc i} 2814.1\AA\ and Fe~{\sc ii} 2813.33 \& 2814.45\AA\ spectral profiles (top panel) and Mg~{\sc ii} 2791.6\AA\ subordinate (middle panel), for flaring pixel y = 119.53\arcsec. The intensity scale is reversed for clarity. The presence of a separate, rapidly evolving strongly redshifted component is obvious for all lines (horizontal dashed lines mark the rest wavelengths). The red diamonds (all panels) indicate the time of the spectra sampled in Fig. \ref{fig:spectraA} and \ref{fig:spectraB}. Bottom panel: space-time plot from a subset of 81 pixels along the IRIS slit (see Figure \ref{fig:irisaia} or Paper I for context). The total fitted intensity of the Fe~{\sc ii}~2814.45\AA\ line is shown in the inverted color table, red diamonds mark the position and times sampled above, while blue diamonds correspond to the spectra in Figures \ref{fig:spectraA_16} to \ref{fig:spectraB_52}.}
 \label{fig:ltspec}
\end{figure}

\begin{figure*}
  \centering
  \includegraphics[width=14cm]{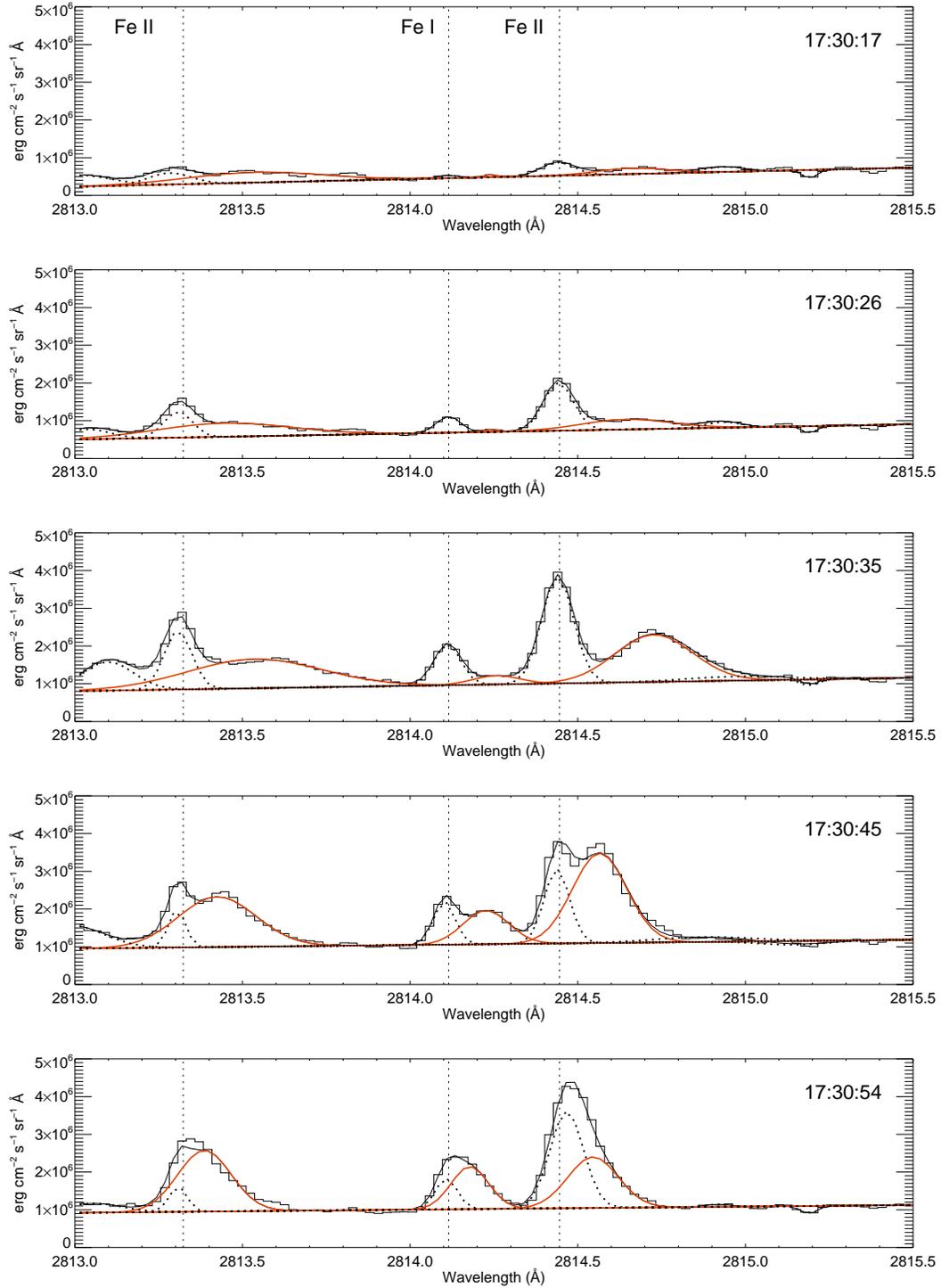}
   \caption{Temporal evolution of the spectra in the Fe~{\sc ii} window for multiple chromospheric lines (black, stepped lines), for the y=119.53\arcsec\ slit pixel (see markings on Figure \ref{fig:irisaia}). The dotted vertical line represents the rest wavelength for each line (labelled in the top panel). Dashed black lines represent the Gaussian fits for the stationary component; red lines the fit for the red-shifted components, and the total fit is given as a black solid line. Note that the Gaussian fits are displayed at the native resolution of the IRIS spectra. Time runs vertically, corresponding to the red diamonds on Figure \ref{fig:ltspec}  (successive exposures at the overall 9.4 s cadence; 2.4s exposure). The intensity scale is kept constant, to better appreciate the variations in brightness during the flare. No subtraction of pre-flare background has been performed.}
  \label{fig:spectraA}
\end{figure*}

\begin{figure*}
  \centering
  \includegraphics[width=16cm]{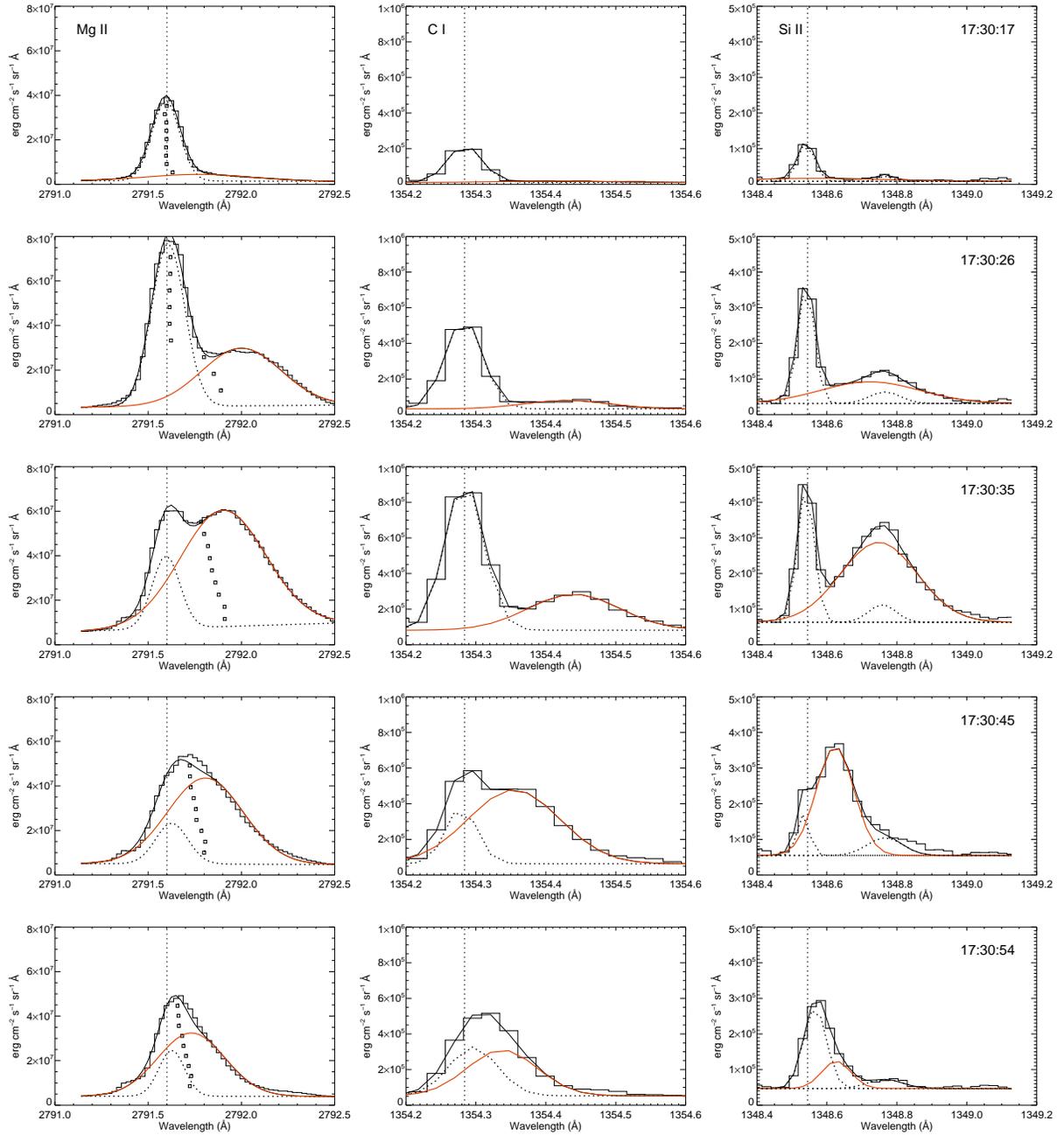}
   \caption{Spectra for the same spatial pixel as described in Figure \ref{fig:spectraA} but for the Mg~{\sc ii}, C~{\sc i}, and Si~{\sc ii} lines. Note that a 2.4s exposure was used for Fe~{\sc ii} and Mg~{\sc ii}, and a 8s one for Si~{\sc ii} and C~{\sc i}. Again, the intensity scale is kept constant for each line. Squares mark the bisectors of the observed Mg~{\sc ii} spectra, at 10\% intensity increments (see Figure \ref{fig:bisec_comp} and the discussion in Appendix \ref{sec:appendixC}.)}
  \label{fig:spectraB}
\end{figure*}

The presence of ``satellite'' red-wing components is clearly observed in multiple chromospheric lines, as vividly demonstrated in 
Figure \ref{fig:ltspec}. In the figure we compare the temporal evolution of  Fe~{\sc i} 2814.11\AA, Fe~{\sc ii} 2813.33\AA\ \& 2814.45\AA, and Mg~{\sc ii} 2791.6\AA~spectra, acquired in one pixel near the brightest part of the ribbon (y $\simeq$ 119\arcsec, cf. e.g. Figures \ref{fig:irisaia} and 
\ref{fig:difference}). After a small rise in intensity shortly after 17:30~UT,  a strong impulsive burst is observed at the lines' undisturbed position, accompanied by a noticeable increase of the continuum level around the Fe lines (the variation of the continuum around the Mg~{\sc ii} 2791.6\AA\ line cannot be clearly appreciated within the small spectral range displayed). At the same time,  a separate component appears far in the red wing of all the lines, and rapidly migrates back towards the lines' centroid over the next 4-5 time steps, i.e. in less than a minute. 

The top panel of Figure \ref{fig:ltspec} shows the Fe~{\sc ii} and Fe~{\sc i} lines, their behaviour essentially identical. The same pattern
 occurs for the Mg~{\sc ii} subordinate line (middle panel), and is observed with varying clarity in many other pixels and lines (see Figures \ref{fig:spectraA}, \ref{fig:spectraB}, and Appendix \ref{sec:appendixA}), so the effect is certainly not due to unidentified lines. Common to all the spectral lines shown in Figure \ref{fig:ltspec} is also a brief ($\sim$ 30 s) and weak (few km s$^{-1}$) blue-ward ``rebound'' occurring around 17:31:30~UT once the red-shifted component has fully migrated back to the rest wavelength. This will be commented upon in Sect. \ref{sec:conclusion}.

\subsection{Condensation: Spectral Fits}\label{sec:chrom_spectra}
In the following we identify the Fe~{\sc i} 2814.11\AA, Fe~{\sc ii} 2813.3 and 2814.45\AA, Mg~{\sc ii} 2791.6\AA, C~{\sc i}~1354.284\AA\ and Si~{\sc ii}~1348.545\AA\ lines as excellent diagnostics of flare chromospheric dynamics: while their intensity is much enhanced during the flare, these lines remain relatively narrow and unsaturated, and can be fairly well isolated from spectral blends and the general background.

Using the pixel at y=119.53\arcsec\ (marked in red diamonds on the bottom panel in Figure \ref{fig:ltspec}) as a particularly clear example, we show in Figure \ref{fig:spectraA} the spectra of multiple chromospheric lines in the Fe~{\sc ii} and Fe~{\sc ii} window, obtained at 5 consecutive time steps (black stepped lines). Fig. \ref{fig:spectraB} shows the spectra of the Mg~{\sc ii},  C~{\sc i}, and Si~{\sc ii} lines in the same pixel, and times. The times in the panels of Figs. \ref{fig:spectraA} and \ref{fig:spectraB} correspond to the red diamonds of Figure \ref{fig:ltspec}, and the dotted black line indicates the nominal rest wavelength of each line. The wavelength scale was established by fitting the photospheric Ni~{\sc i} 2799.474\AA\ line over a 3 hour period before the flare, to remove the residual orbital oscillation as described in the IRIS technical note ITN20. We also use the procedure described in IRIS Technical Note ITN24  \footnote{The IRIS technical notes can be found at http://iris.lmsal.com/documents.html.} to convert the raw DN output to intensity, as done e.g. by \citet{2016ApJ...816...88K} and \citet{2017ApJ...836...12K}. No subtraction of pre-flare background has been performed.

In Figs. \ref{fig:spectraA} and \ref{fig:spectraB} we see that all lines show a strong intensity increase at 17:30:26 (second row), with a concomitant, broad satellite component appearing at high red-shifts. The shifted component is weaker at first detection, but in the next two time steps its intensity becomes comparable with the main peak, which remains mostly stationary. While increasing in intensity, the red-shifted component rapidly migrates towards the rest wavelength, so that by 17:30:54 (only 37 s from the ``pre-flare'' time) the general appearance is that of a single, broad, asymmetric line. It is worth noticing that the red-shifted component is more prominent in the Mg~{\sc ii} spectra (Fig. \ref{fig:spectraB}), and can often be unambiguously detected in them one time-step ahead of the other lines (cf. also the figures in Appendix \ref{sec:appendixA}). This is probably due to the much larger opacity of this line with respect to the other chromospheric signatures.

Most of the flaring pixels display a behavior similar to that shown in Figs. \ref{fig:spectraA} and Fig. \ref{fig:spectraB}, with spectral profiles very asymmetrical during their early activation, due to the presence of the satellite component, and a clear continuum enhancement from the pre-flare value, best observed in the Fe~{\sc ii} window. As an example, in Appendix \ref{sec:appendixA} we show corresponding Figures for two additional pixels (y $\sim$ 117\arcsec, and y$\sim$ 123\arcsec~ along the slit); these same pixels are identified with the blue diamonds in Fig. \ref{fig:ltspec}.

To characterize the dynamics of the condensation, we have thus fitted the profiles of each flaring pixel
with a pair of Gaussian profiles. As the shifted component is so prominent, a Gaussian fit should provide a better estimate of the redshift over the traditional bisector method that we employed in Paper I, which we find can slightly underestimate the maximum shift (in Appendix \ref{sec:appendixC} we make a side by side comparison of both techniques.) The weak spectral lines such as Fe~{\sc ii} and Fe~{\sc i} are most likely optically thin \citep[see discussion in][and Sect. \ref{sec:models} below]{2017ApJ...836...12K}, so the use of a Gaussian approximation is further justified. For stronger lines, like the Mg~{\sc ii} 2791.6\AA, we follow the original work of \citet{1984SoPh...93..105I} in assuming that the flare emission originates
mostly from a separate, optically thin slab of (moving) plasma overlaying the stationary atmosphere.
In Figs. \ref{fig:spectraA} and \ref{fig:spectraB}, the quasi-stationary component of each spectral line is shown with black dotted lines, while the red-shifted component is displayed in red. The final overall fit is given as a thin black line.
We constrain the central component in all lines to remain within $\pm$~2 km/s of the rest wavelength, allowing the second shifted component to return as far as possible towards the line rest wavelength, and track its complete velocity evolution. At times, the spectral range in proximity of Fe~{\sc ii} 2813.3 \AA\ can be subject to some ambiguities due to the presence of other, much weaker Fe~{\sc i} and Fe~{\sc ii} lines, but the fits are satisfactory overall. The same problem occasionally affects the fits for the Si~{\sc ii} line (Fig. \ref{fig:spectraB}), due to an unidentified weak line that is visible in the later exposures around 1348.75\AA. Its profile remains low in the red wing of the Si~{\sc ii} line, so we constrain the maximum intensity and position to avoid double fitting the shifted, satellite Si~{\sc ii} line \citep[cf. e.g. the list of adjacent spectral lines in][]{1986ApJS...61..801S}. The stationary component of Si~{\sc ii} also has a tendency to peak just blue-ward of the rest wavelength during the rise phase. The profile is itself slightly asymmetric, perhaps indicating an unidentified blend not listed in the \cite{1986ApJS...61..801S} atlas or a difference in the formation height of the ion compared to the other lines shown. Aside from these small uncertainties, the general behavior remains very clear in all lines.

Lastly, we notice that the width of the red-shifted components appears to be very large with respect to that of the stationary component, and to the pre-flare situation in general. While some broadening could be expected due to the long exposure times coupled with the rapid motion of the component (particularly in the FUV), it is plausible that most of the excess width could be created by flare-induced turbulent motions in the chromosphere \citep{2011ApJ...740...70M, 2015ApJ...804...56R}; the fact that the width of the red-shifted component seem to decrease with time (cf. the top and bottom panels in Fig. \ref{fig:spectraB} might support this interpretation.

\subsection{Condensation: Time Evolution}\label{sec:chrom_time}

We now focus on the dynamics displayed by multiple flaring pixels, in particular those comprising the ribbon between the
y=[114$''$,127$''$] positions in Fig. \ref{fig:ltspec}. These are the same pixels analyzed in Paper I.

Using the fits described in Section \ref{sec:chrom_spectra}, we derive the centroid of the second, shifted Gaussian component for each line, pixel and time step. The velocity scale for each line is obtained with respect to the rest wavelength shown in Figs. \ref{fig:spectraA} and \ref{fig:spectraB}. To remove fits where the velocity determination could be biased by nearby lines or low signal, we discarded the fits when the total intensity of the shifted component was below 30\% of the maximum reached in any given pixel. This also removes some of the influence of the instrumental point spread function (PSF) described in Appendix \ref{sec:appendixB}.

\begin{figure}
  \centering
  \includegraphics[width=9cm]{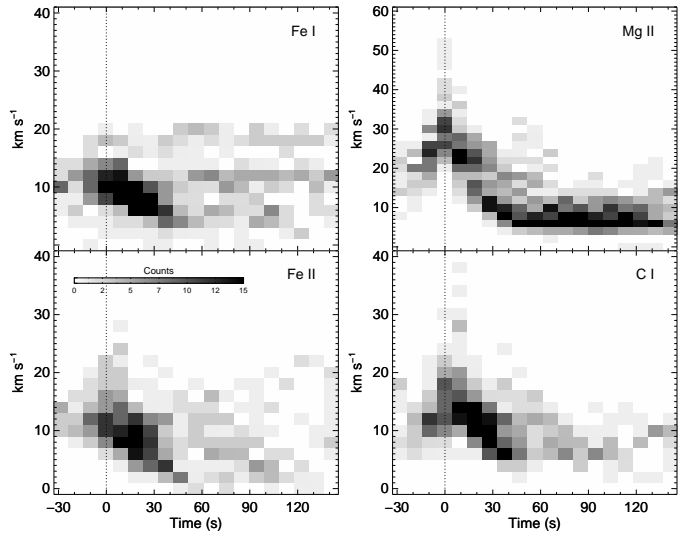}
   
   \caption{Superposed epoch analysis of chromospheric condensation flows derived from the second component of the IRIS lines' fits. The
   velocity curve for each pixel is aligned in time to the occurrence of maximum centroid shift for the second component of Mg~{\sc ii} at that pixel. The velocity indicates down-flow motions and the strength of the colortable dictates the number of instances within a 2 km/s velocity interval for each time step.}
  \label{fig:condensation}
\end{figure}

To find commonalities in the flow evolution of the different pixels and lines, in analogy to Paper I we perform a superposed-epoch 
analysis, aligning the velocity-time curve for each pixel and spectral line to a common zero time. The latter is defined 
as the time at which maximum shift of the second component in Mg~{\sc ii} is reached for any given pixel. Figure \ref{fig:condensation} displays the resulting curves. For each time step (given by the cadence of the observations, 9.4s) the number of pixels within a 2 km/s interval is represented by the gray scale. The bar in the figure gives the number of pixels within each time-velocity bin.

From Figure \ref{fig:condensation} we can immediately confirm that the behaviour described in Sect. \ref{sec:chrom_spectra}
is common to the majority of the flaring pixels in the ribbon sampled by the IRIS slit.
While clearest for the Mg~{\sc ii} line, the evolution of the red-shifted component for all considered pixels and spectral lines falls within a fairly narrow envelope, starting at a maximum redshift of
up to 50 km s$^{-1}$, and rapidly decaying to a rest velocity within $30-60$ seconds. This confirms and expands our 
previous findings (cf. Fig. 5 in Paper I), that 
the ``satellite'' component of chromospheric lines provides a clear picture of the condensation occurring within each flaring pixel 
at the early stages of the flare. The overall decay curve is essentially identical in all lines, even if the spectra in the 
IRIS FUV and NUV ranges are acquired with different exposure times (8 s and 2.4 s, respectively). 
Thus, it seems that the 9.4 s cadence of the observations is rapid enough to preserve the basic properties of the condensation evolution \citep{1989ApJ...346.1019F, 2017ApJ...836...12K}, although higher cadence data could provide further insight. 

After $\sim$ 60 s from the peak velocity, the fits for the red-shifted component in most lines become ambiguous, and a large scatter 
is thus observed in the time-velocity plot. For the Mg~{\sc ii} line, a fit is instead possible for a longer period, and a 
fairly constant condensation velocity of few km s$^{-1}$  is observed for several minutes (not fully shown in Figure 
\ref{fig:condensation}).

Finally, from Fig. \ref{fig:condensation} we see that a ``build up phase'' of the flows is apparent in Mg~{\sc ii}, and possibly in other lines, 1-2 time steps before the peak. This corresponds to the presence of a weak red-shifted component for the Mg~{\sc ii} line, that can be observed earlier than for other lines (cf. the top panels of Fig. \ref{fig:spectraB}, \ref{fig:spectraB_16} and \ref{fig:spectraB_52}). It is however difficult to say whether this is due to a differences in opacity between the lines or to some artifact of the PSF (see Appendix \ref{sec:appendixB}).

\section{HXR spectra} \label{sec:fermi}

Modern radiative-hydrodynamical models of the flaring lower solar atmosphere like the flare RADYN code 
\citep{2015ApJ...809..104A} usually assume that the flare energy is transported from the corona by means of a beam of accelerated, mildly relativistic electrons. The parameters characterizing the electrons beam can be derived by analyzing the HXR spectra originating from the flaring 
atmosphere, assuming that the bremsstrahlung emission derives from the impact of such beam onto the dense chromosphere
\citep[e.g.][]{1971SoPh...18..489B}.

To this end, we analyzed the HXR data acquired by the
Fermi/Gamma-Ray Burst Monitor (GBM). As mentioned in Section \ref{sec:flare_description}, Fermi fully covered the 
impulsive phase of the flare. GBM continuously observes the whole unocculted sky, 
recording two different data types: CTIME data, with a fine time resolution 
(0.256~s; 64~ms on flare trigger), but a coarse spectral resolution in 8 
energy channels; and CSPEC data, with lower time resolution (4.096~s; 1.024~s 
on flare trigger), and a spectral resolution of 128 energy channels over the 
8~keV - 1~MeV spectral region. In the following we use CSPEC data, acquired with the 
NaI02 detector, which was the most favorably oriented one among the twelve 
sodium iodide (NaI(Tl)) scintillation detectors.

\begin{figure}[!hbt]
  \begin{center}
    \includegraphics[width=8cm]{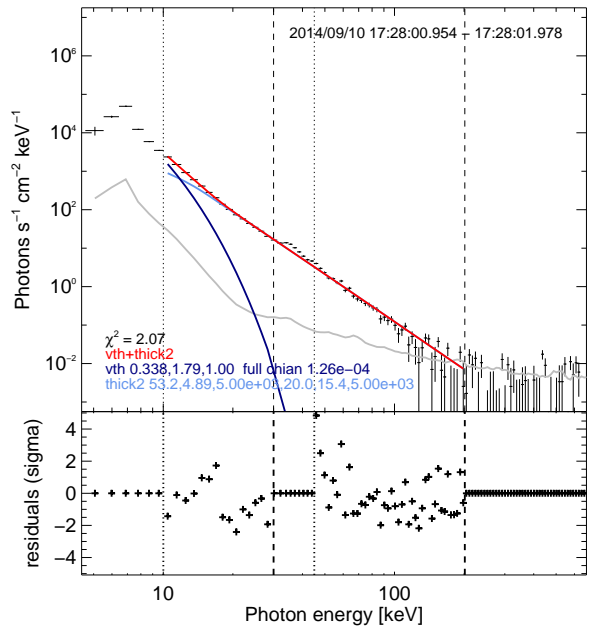}
    \caption{ Top panel: HXR spectrum around the time of flare maximum, from Fermi-CSPEC 
      data (black points with error bars). In gray is the pre-flare background flux averaged between 17:13 and 17:15~UT. The colored curves represent the fit obtained from the OSPEX forward fitting procedure, including a thermal component (dark blue), 
      a single power-law component (light blue), and the combined fit (red).
      Dotted and dashed lines indicate the lower and upper limits of the 
      two energy ranges used for the non-thermal component fit (10--30 keV and 45--200 keV); this is 
      necessary due to calibration issues that affect the intermediate energies (see text).
      Bottom panel: the residuals between the fitted distribution and real data, normalized to
      the 1-$\sigma$ uncertainty for each respective data point. The fitted distribution
      can be considered acceptable as most of the points lie within 2-$\sigma$.
    }
      
    \label{fig:fermi_spectrum}
  \end{center}
\end{figure}

The count rates profiles integrated in four energy bands are shown in Figure~\ref{fig:hxr_intensity} and briefly 
discussed in Section \ref{sec:flare_description}. The flare was so strong that 
the 1~s cadence spectra have a high signal-to-noise ratio (SNR), so we can reliably derive the HXR spectral parameters
at this high temporal resolution.  An example of a single Fermi spectrum acquired during the maximum of the flare is given 
in Figure~\ref{fig:fermi_spectrum}.

Assuming that the bremsstrahlung 
emission originates in a cold thick-target \citep{1971SoPh...18..489B}, 
we deduce the accelerated electrons' energy spectrum as characterized by the total flux of non-thermal electrons, 
a power-law index $\delta$, and a low energy cut-off $E_{\rm c}$ of the 
electron energy distribution. The parameters characterizing the low-energy, 
thermal component are the electron temperature and the emission measure. 

The total non-thermal electron power is calculated using the following 
relation \citep{2011SSRv..159..107H}:

\begin{equation}
   P_{\rm nth}=k_{\rm E}\int_{E_c}^{+\infty}{E_0{\cal{F}}_0(E_0)}dE_0=
   \frac{k_E A}{\delta -2}E_c^{-\delta +2}~{\rm erg~s^{-1}}
\end{equation}

where ${\cal{F}}_0$ is the injected electron beam flux distribution, $E_c$ is 
the low energy cut-off, $\delta$ is the spectral index of the electron energy 
distribution, $k_E$ is the conversion from keV to erg, and $A$ is the 
normalization factor of the electron flux distribution. The total non-thermal 
electron energy input in the flare can be estimated by integrating 
$P_{\rm nth}$ over the flaring time interval.

The data analysis and the fitting procedure have been performed with the OSPEX 
routines of SolarSoft package \citep{2002SoPh..210..165S}. As the Fermi/GBM data  have a known calibration issue at the Iodine K-edge energy of 33.17 keV, resulting in a discontinuity in the response of the NaI detectors in the case of bright bursts \citep{2009ExA....24...47B}, we have conservatively excluded the 30--45 keV range from the fits. This is indicated with the dotted and dashed lines in Fig. \ref{fig:fermi_spectrum}.

We have also opted not to include an albedo component in the fit, although this could potentially be significant in the HXR spectra due to the location of the flare close to disk center \citep{Santangelo:1973SoPh...29..143S}. However, several assumptions that enter the possible albedo contribution, in particular the pitch-angle distribution function of the non-thermal electrons, and the absence of any magnetic mirroring effects \citep{Kontar:2006A&A...446.1157K, Dickson:2013SoPh..284..405D}, are not independently verifiable. The most relevant contribution of the albedo component is expected in the range $\sim$ 15--50 keV \citep{Santangelo:1973SoPh...29..143S,Kontar:2006A&A...446.1157K}, i.e. overlapping the 30--45 keV range that we excluded from the fit due to the instrumental effects mentioned above. We thus feel that our choice is justified, and note that the results of the fit including the albedo contribution would only potentially decrease the total number of non-thermal electrons by a factor of about 3 \citep{Simoes:2013A&A...551A.135S}, well within our final estimate of the overall flux (cf. Sect. \ref{sec:uv_vs_hxr}).




Most of 
the spectra observed during the impulsive phase are well represented by a thermal 
component (dark blue in Figure \ref{fig:fermi_spectrum}) plus a single power-law component (light blue), 
describing the electron 
bremsstrahlung emission in the 10-200~keV range. 
During the strongest HXR 
intensity peaks, however, the spectral fit procedure can at times return somewhat ambiguous results, 
as multiple solutions can be obtained for the same spectra. A major cause of this ambiguity is the balance between thermal and non-thermal components that can be assigned in the final spectral characterization. For this reason our approach has been to derive the electron temperature from simultaneous GOES observations \citep[see][]{2005SoPh..227..231W}, thus better constraining the thermal component. 
In general, the $\delta$ spectral index is well characterized in the fits, while
the estimate of the power contained in the non-thermal electrons strongly depends on the 
total flux of non-thermal electrons and the low energy cut-off value. Better constraining the temperature in this case provides a more realistic estimate of the low energy cut-off, critical for modeling (see Section \ref{sec:models}).

\begin{figure}[!hbt]
  \begin{center}
    \includegraphics[width=8cm]{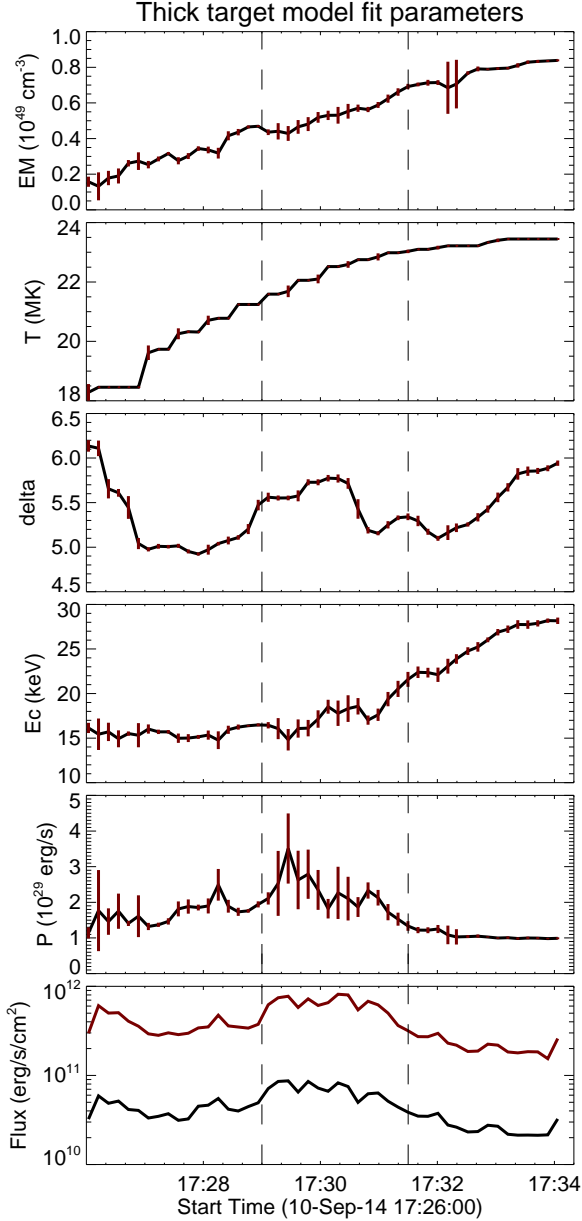}
    \caption{Time profile of the HXR spectral parameters obtained from the fits of Fermi spectra at 
    1 s cadence.  Values shown are averaged over 10 s; vertical brown bars represent the variations over such intervals (some bars are so small to
    be hidden in the plot). From top to bottom:  emission measure and electron temperature of the thermal component; spectral index $\delta$, cutoff energy and total power of the non-thermal electrons' energy  distribution; non-thermal electrons' flux. The latter requires an estimate of the area involved in the 
 HXR emission, as described in Sect. \ref{sec:uv_areas}.}
    \label{fig:hxr_fit_parameters}
  \end{center}
\end{figure}

The evolution of the model parameters derived from the fits are shown in
Figure~\ref{fig:hxr_fit_parameters} for an 8 minute interval during the flare's
impulsive phase. 
Given that the IRIS data are acquired with a 9.4~s 
cadence, after fitting all the spectra at their original time resolution of 1 s we calculated the 
mean and the rms of the spectral parameters over 10 s intervals, assuming that the dispersion of
their values mostly represents the uncertainties of the fitting procedure rather than a real
evolution of the spectra within 10 s. The short vertical bars in Fig. ~\ref{fig:hxr_fit_parameters} indicate the range of variation 
for each parameter in such intervals.
The dashed lines, also outlined in Fig. \ref{fig:hxr_intensity},
frame the period of maximum UV emission and condensation flows for the 
flaring pixels sampled by the IRIS slit.
Figure~\ref{fig:hxr_fit_parameters} reveals fairly hard spectra with $\delta \approx 5$, in 
particular during the first intensity peak phase around 17:27-17:28 UT (cf. Fig. \ref{fig:hxr_intensity}). 
The non-thermal electron power 
however reaches its maximum of $P_{nth} \approx 2-3 \times 10^{29}$ erg s$^{-1}$ 
within the 17:29-17:31 interval, which corresponds to the maximum emission in the IRIS channels.  This is
also the interval when maximum flux is inferred, as described in the next section.

Finally, by using the derived values of non-thermal electron power integrated over the whole
flare duration (from 17:20:16 to 17:51:00, i.e. significantly longer than the interval shown in Fig. \ref{fig:hxr_fit_parameters}),
we estimate a total energy content in the accelerated electrons of about 
$10^{32}~{\rm erg}$. This is broadly consistent with values reported in the literature for large eruptive flares, e.g. \citet{2012ApJ...759...71E}.

\section{Area and duration of energy release episodes}\label{sec:uv_areas}

The spatially unresolved Fermi observations are ill-suited to estimate two parameters crucial to a full description of the electrons' beam: 
the energy flux (erg s$^{-1}$ cm$^{-2}$) impinging on the chromosphere,
which depends on the (time-varying) area involved, and the duration of such energy input in any given area. 
These values can be derived from RHESSI combined imaging and spectroscopy when available; 
even so, they often suffer from large uncertainties due to poor spatial and temporal resolution, or noise in the data. 
For this reason, it has become customary to estimate the area interested by electron precipitation 
directly from the size of optical and UV 
chromospheric brightnenings \citep[e.g.][]{2011ApJ...739...96K,2015ApJ...813..125K,2015ApJ...798..107K,2017ApJ...836...12K}. 

The duration of each energy deposition episode is instead estimated using various techniques, including the rising time of individual
kernels' UV brightness \citep{2010ApJ...725..319Q, 2012ApJ...752..124Q}, the duration of ``single'' bursts in the HXR (or derivative of soft X-ray) 
curves \citep{1995A&A...299..611C,1996A&A...306..625C,2016ApJ...827...38R}, or the comparison of coronal diagnostics
with multi-thread modeling employing the duration as a free parameter \citep{2006ApJ...637..522W}. In the following we attempt to provide realistic estimates for these two beam parameters for the case of SOL2014-09-10T17:45 by using the concomitant Fermi, IRIS and AIA observations.

\subsection{HXR energy flux}\label{sec:uv_vs_hxr}

The close temporal correlation of HXR and UV emission during flares has long been recognized \citep[see e.g.][]{1981ApJ...248L..39C,1988ApJ...330..480C}, and is generally understood in terms of both signatures being produced by the impulsive heating of the lower atmosphere from precipitating non-thermal electrons.
Observations with higher spatial resolution have however clarified how the global UV emission curves result from
the staggered occurrence of multiple flaring kernels, each one displaying a similar evolution of UV emission, with a rapidly rising phase and a much longer cooling decay  \citep[][see also Fig. \ref{fig:ribbon} below]{2010ApJ...725..319Q, 2012ApJ...752..124Q}. Details of the actual energy deposition are encoded mainly in the rising phase of such ``individual'' curves. 

\begin{figure}
  \centering
  \includegraphics[width=10cm]{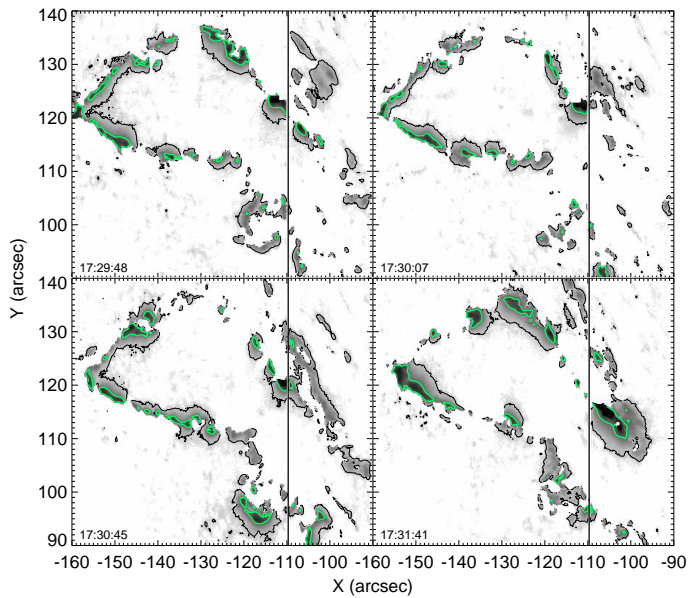}
   \caption{Positive running difference of SJI 2796\AA\ images for four representative times during the flare development (inverted, log
   intensity scale). Green and black contours represent 130 and 9 DN intensity change, respectively. A reduced FOV surrounding the most intense part of the flare ribbon is shown (cf. Fig. \ref{fig:irisaia}); the IRIS slit is visible as the black line.}
  \label{fig:difference}
\end{figure}

We exploit this property of the UV emission by computing the running difference of the IRIS 2796\AA\ SJIs at their original
cadence of 19 s, with the idea that every flaring area undergoing heating would be clearly recognized in such difference images (the 1400\AA\ images could not be used as overly saturated). Figure \ref{fig:difference} shows the difference images for four representative times during the flare development; only the positive intensity changes from the prior image are displayed using an inverted color table with logarithmic scaling; a threshold is used to remove the background noise. Every structure visible in the images represents an area which is either newly activated, or still in the rising phase of the UV curve -- hence still experiencing energy input at that given moment. Two contour levels are shown in the difference images: the most intense regions are defined at a change of more than 130 DN, containing between 40-50\% of the total intensity change in each image, and displayed in green, while the weaker enhancements of the ribbon chromosphere are shown in black at a level of 9~DN, approximately 5 times the noise --- as determined from the average counts in an area free of ribbon emission in the bottom-left corner.

For any given time, we computed the area of the flaring sources from the total number of new pixels within 
each of these contours. Note that a further correction is required to take into account the small part of the ribbon area that is not 
imaged by IRIS SJIs (see Figure \ref{fig:irisaia}); to this end we introduce a factor determined from the ribbon area at the 2000 DN 
level in the AIA 1600\AA\ image acquired closest in time. In the 17:20 and 17:50~UT interval this factor varies between 1.0 and 1.2.

\begin{figure}
  \centering
  \includegraphics[width=9cm]{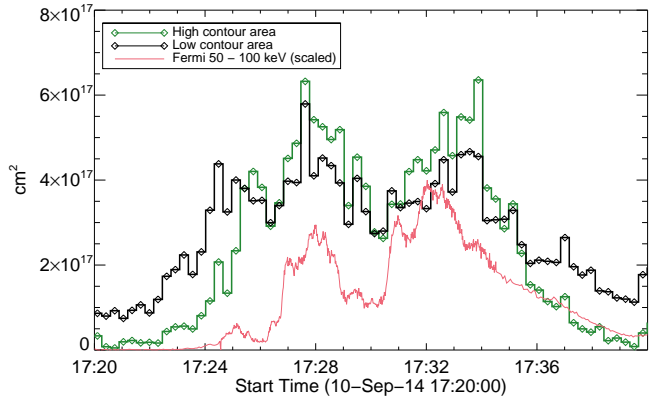}
   \caption{Newly activated kernel area in comparison with the Fermi 50-100 keV lightcurve (normalized and scaled to the plot window). The area was taken from 2796\AA\ SJIs at the level of the green and black contours in Figure \ref{fig:difference} and includes a correction for the missing ribbon area seen in the AIA 1600\AA\ images.}
  \label{fig:uv_hxr}
\end{figure}

A strong correlation exists between the ``active'' ribbon area and the HXR emission. This is apparent in Figure \ref{fig:uv_hxr}, where the area (in cm$^2$) calculated from the strongest contours of Fig. \ref{fig:difference} is plotted alongside the 50-100~keV emission from Fermi. In particular, during the central portion of the impulsive phase of the flare (17:26:30 -- 17:32:30 UT) most HXR bursts correspond to new peaks in
the area curve. The relationship is tight in the temporal sense, but not in amplitude:
at times relatively modest enhancements in the HXR emission might correspond to large increases in area (cf. the interval 17:29--17:30 UT), while at others the opposite occurs (interval 17:30:30--17:31:30 UT). The main point of Fig. 
\ref{fig:uv_hxr} is then a confirmation of the results of \citet{2010ApJ...725..319Q, 2012ApJ...752..124Q} reported above, i.e. that
the UV emission in any given area can be meaningfully correlated with the HXR signal, as a tell-tale of accelerated electrons impacting 
the chromosphere, only during its rising phase. 

An average beam energy flux at any given time during the flare can be estimated using the flaring area values calculated above.
Given the temporal evolution of the power in the non-thermal electrons as derived in the previous Section, and the two area limits as described above, we find the lower and upper limit to the average flux to be on the order of $\approx 10^{11}-10^{12}~{\rm ergs~cm^{-2}~s^{-1}}$, shown in the bottom panel of Figure \ref{fig:hxr_fit_parameters}. These are rather high fluxes, at the limit of what current radiative hydro-dynamical simulations assume
for typical solar flaring conditions. Yet, values of fluxes of the order of several $10^{11} {\rm ergs~cm^{-2}~s^{-1}}$
are also derived in recent papers that utilize high spatial resolution UV and optical data \citep{2011ApJ...739...96K,2015A&A...578A..72K,2015ApJ...813..125K,2017ApJ...836...12K,2019ApJ...878..135K}.

From the curves of Fig. \ref{fig:uv_hxr} we realize that there is no obvious reason for the HXR spectrum to be self-similar in each flaring kernel at any given time, so injection energy rates could be widely different than the spatial averages just derived. However, during the interval most relevant for our analysis (17:29:48 - 17:31:41~UT), the flaring area directly sampled by the IRIS slit was the brightest UV emitter and, as mentioned in Sect. \ref{sec:flare_description}, its intensity accounted for 15-20\% of the total 2796 \AA\ SJI counts, while accounting for only $\sim$1\% of the total area. For this reason we think that 
a flux value of few $\times 10^{11} ~{\rm ergs~cm^{-2}~s^{-1}}$  is a reasonable assumption, although higher fluxes cannot be
excluded.

\subsection{Duration of energy release}\label{sec:duration}

Both Figure \ref{fig:difference} and Figure \ref{fig:uv_hxr} highlight how each HXR bursts 
might pertain to multiple, separate locations, that rapidly evolve during the flare development. 
Since the duration of heating in any one of these ribbon pixels can not be directly determined
without spatially resolved HXR imaging, we use the UV light-curve as a proxy, as mentioned above.

\begin{figure}
  \centering
  \includegraphics[width=9cm]{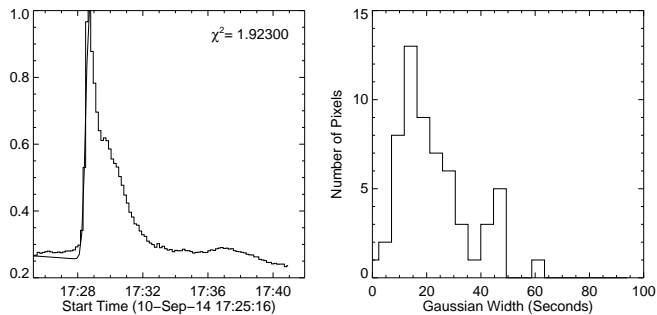}
   \caption{Left panel: example lightcurve of normalised Fe~{\sc ii} 2814.45\AA\ intensity (stepped line) with a Gaussian fit (smooth line) to the data prior to the peak; the same pixel as in Figure \ref{fig:spectraA}. Right panel: histogram of fitted Gaussian widths in seconds, for fits with reduced $\chi^{2} \geq 0.5~{\rm and}~\chi^{2} \leq 5$.}
  \label{fig:ribbon}
\end{figure}

Shown in Figure \ref{fig:ribbon} is an example of a single pixel light-curve of the total Fe~{\sc ii} 2814.45\AA\ line intensity (a sum of the spectrum between 2814.24\AA\ and 2815.0\AA). Following \citet{2012ApJ...752..124Q}, we fit the rise side of each pixel's 
light curve, from background to peak intensity, with a Gaussian profile, and assume its full Gaussian width as the heating time for that pixel.

Most of the 81 flaring pixels return a good fit, but to determine an average heating time we consider only pixels with a reduced $\chi^{2}$ between 0.5 and 5. A histogram of the relevant Gaussian widths is shown, in units of seconds, in the right hand panel of Figure \ref{fig:ribbon}. The median width is $\sim$22 s, with the peak in the histogram around 15 seconds. We will thus assume a heating duration of $\sim$20 s in our modeling. This is consistent with the recent work of \citet{2016ApJ...827...38R} and \citet{2017ApJ...836...12K}, that also analyze high resolution IRIS spectra of an X-class flare.

\section{Model} \label{sec:models}
\subsection{Initial set up}\label{sec:models_setup}

We modeled the impulsive phase of the SOL2014-09-10T17:45 event using the RADYN flare code \citep{1997ApJ...481..500C,2015ApJ...809..104A}. The code solves the time-independent Fokker-Planck (F-P) equation for a prescribed injected particle beam distribution, given the charge and mass of the beam particle, the initial pitch-angle distribution in the downward hemisphere, the power-law index, the low-energy cutoff value, and the energy flux density at the top of a model 1D magnetic loop consisting of a photosphere, chromosphere, and corona.

Following the analysis of Sects. \ref{sec:fermi} and \ref{sec:uv_vs_hxr}, we use a flare model with the following electron beam parameters: $\delta$=5, E$_c$ = 15 keV, and energy flux of $10^{11} ~{\rm ergs~cm^{-2}~s^{-1}}$ (F11). The energy input lasted 20s. The model QS.SL.HT of \citet{2015ApJ...809..104A} was used for the pre-flare atmosphere, as it was closest to the observed plage environment. Several improvements have been made to the RADYN flare code since \cite{2015ApJ...809..104A}, which are worth noting here \citep[they will be described further in][in prep]{2020_allred_inprep}. The hydrogen broadening from \cite{2017ApJ...837..125K} and \cite{2009ApJ...696.1755T} have been included in the dynamic simulations 
(Kowalski et al. in prep.).
The QS.SL.HT model was relaxed with this new hydrogen broadening, and we choose to use the X-ray back-heating formulation from \cite{2005ApJ...630..573A} for these models; 
the resulting pre-flare apex temperature is 1.8 MK, with electron density $5.1 \times 10^{9} ~{\rm ~cm^{-3}}$. Finally, we used a new version of the F-P solver,
which gives a moderately smoother electron beam energy deposition profile over height in the upper chromosphere.

To properly compare the model results with the observations, we then synthesized the Fe~{\sc ii} 2814.45\AA~line at different times within the evolution of the flare. As shown by \citet{2017ApJ...836...12K,2019ApJ...878..135K}, this line is an important diagnostic in flares because the intensity originates from a similar temperature range, with a broad peak around $T\sim12,000$ K, as the hydrogen Balmer bound-free radiation that dominates the IRIS NUV spectral range. Further, the line can be efficiently synthesized in LTE, by using snapshots of the non-equilibrium ionization, electron density and velocity stratification from RADYN \citep{2017ApJ...836...12K}, and non-LTE temperature. Stronger, optically thick lines, including the Mg~{\sc ii} triplet lines, require a more careful treatment, which is beyond the scope of this paper \citep[cf. e.g.,][]{2019ApJ...879...19Z}.

The calculation of Fe~{\sc ii} profiles remains the same as in \cite{2017ApJ...836...12K}, except that we account for the upper photospheric ($z<150$ km) Mg~{\sc ii} wing opacity in LTE. This provides a more accurate contrast of the chromospheric flare radiation against the upper photospheric (non-flaring) emission. This is important for assessing faint line emission from the chromosphere, such as the Fe~{\sc ii} line. \cite{2019ApJ...879...19Z} show that even a 30$\times$ increase in the expected Stark broadening of the Mg~{\sc ii} line produces very little wing emission at wavelengths $>2806$\AA\ in the flare chromosphere \citep[see Figure 4(a) in][]{2019ApJ...878..135K}, thus we expect minimal influence in the Fe~{\sc ii} window. Finally, for simplicity, a micro-turbulence parameter was not included in the broadening prescription of the Fe~{\sc ii} 2814.45\AA~line.

\begin{figure*}
  \centering
  \includegraphics[width=\textwidth,height=12cm]{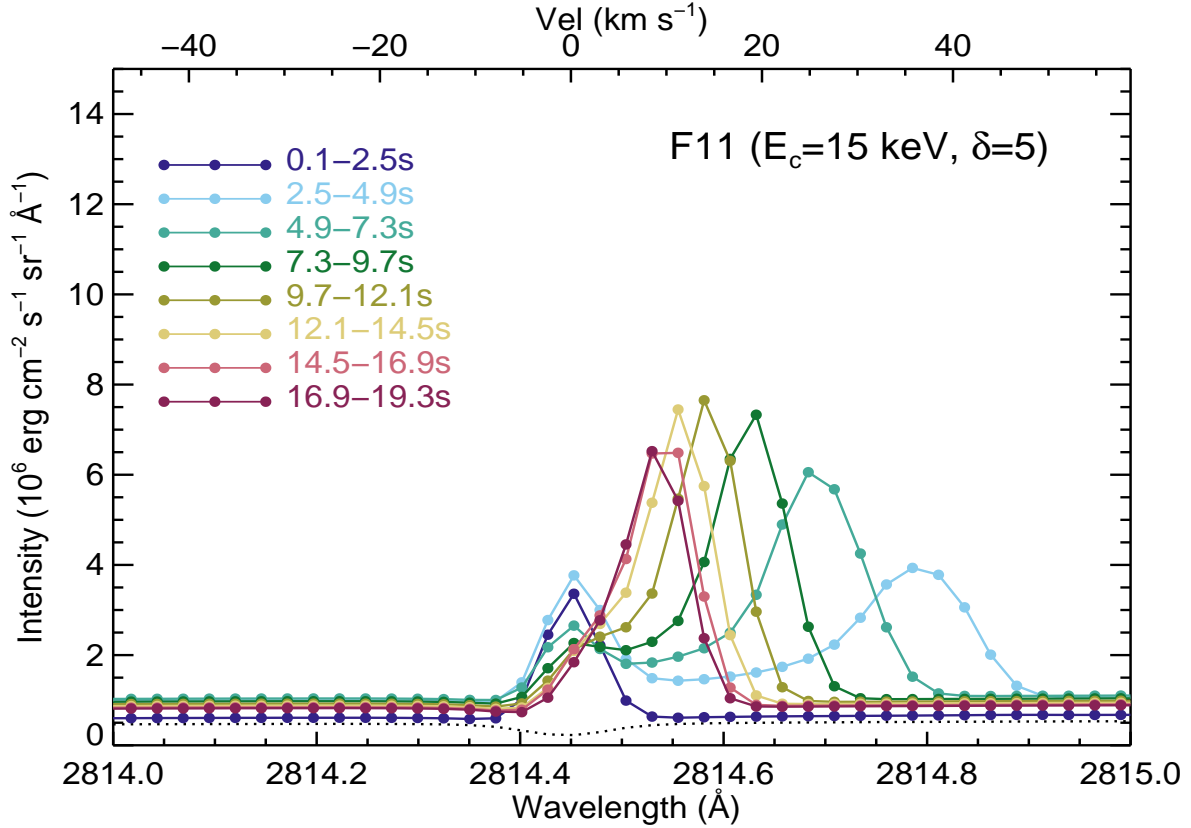}
   \caption{Synthetic spectra of Fe~{\sc ii} 2814.45\AA~obtained at different times during the 
   evolution of the model flare. The profiles are averaged over the duration of the single IRIS exposure, i.e. 2.4 s. The black dotted line represents the pre-flare profile.}
  \label{fig:syntheticspectra}
\end{figure*}

\subsection{Model results}\label{sec:models_results}

Using the prescriptions described above, we calculated the spectra of the Fe~{\sc ii}  2814.45\AA~line at every 0.1~s in the simulations, and averaged them over the exposure time of the NUV IRIS data during the flare, i.e. 2.4 s.
The resulting spectra are displayed in Fig. \ref{fig:syntheticspectra} for the whole duration of the impulsive phase, i.e. 20 s. The pre-flare intensity is not subtracted and is shown as a thin, black dotted line. During the whole evolution, the Fe~{\sc ii} 2814.45\AA~line (both components, see below) remains optically thin, as was the case for the flare described in \citet{2017ApJ...836...12K}.

Several important observational features appear to be reproduced by the model. First, the modeled continuum intensity varies from a pre-flare 
value of $\sim0.5$  to a maximum value of $\sim 1\times 10^6$ erg cm$^{-2}$s$^{-1}$sr$^{-1}$\AA$^{-1}$ at 10--20 s 
into the flare; this is in good agreement with the observed continuum value and its enhancement (cf. Fig. \ref{fig:spectraA} and Figs. \ref{fig:spectraA_16} and \ref{fig:spectraA_52} in Appendix \ref{sec:appendixA}). 
Second, the modeled Fe~{\sc ii} 2814.45\AA~line goes into emission almost instantaneously as the flare develops, and reaches a maximum intensity of $\sim 4\times 10^6$ erg cm$^{-2}$s$^{-1}$sr$^{-1}$\AA$^{-1}$, which is again consistent with the observed intensity values. Finally, the modeled spectra show the rapid development of a strong red-shifted additional component to the line, corresponding to a velocity of $\sim 40$ km s$^{-1}$, that appears just a few seconds into the flare. This second component evolves very rapidly, and becomes dominant with respect to the stationary component (still 
in emission) while decelerating at the same time: only 10 s after its first appearance, its velocity had decreased to $\sim 10$ km s$^{-1}$. 

The existence of a second, red-shifted spectral component, and its dynamical evolution well
resemble the behavior displayed by the observed IRIS spectral lines (Figs. \ref{fig:ltspec} and \ref{fig:spectraA}). A similar dynamical evolution of the Fe~{\sc ii} line has been discussed by \citet{2017ApJ...836...12K} for the case of a stronger flare (5F11, $\delta$=4.2), and explained with the development of a strong chromospheric condensation (see Sect. \ref{sec:discussion} below); however, our data allows a more detailed comparison with the whole condensation evolution, thanks to their high cadence and the availability of multiple flaring kernels observed continuously from the earliest stages.
 
Some discrepancies remain between the data and the modeled spectra, that might point to necessary modifications of the models and/or the spectral synthesis. From the dynamical point of view, the evolution of the synthetic line intensity, especially that of the red-shifted component, appears sensibly faster than in the observations. In Fig. \ref{fig:syntheticspectra} the red-shifted component appears just a few seconds after the heating starts, with an intensity comparable to that of the stationary component. The data, on the other hand, show a more gradual enhancement of both the stationary and red-shifted components, with the latter matching the stationary component intensity only over $\sim$ 30 s. The deceleration of the red-shifted component in the data also appears slower than in the model, with a typical decay time of  $\sim$ 30~s (Fig. \ref{fig:condensation}) vs. the $\sim$ 10 s of the simulations. Finally, the data show a build-up (over a period of 10-20 s) towards the maximum red-shift, at odds with the instantaneous appearance of the second component in the simulated spectral profiles; this effect is better visible for the stronger lines like Mg~{\sc ii} and C~{\sc i}  (Fig. \ref{fig:condensation}).  This could merit further investigation in the context of flare precursors that are often reported \citep[e.g.][]{2017ApJ...840..116B}. 

The intensity of the second component relative to the stationary one also appears larger in the synthesized lines with 
respect to the observations (cf. Fig. \ref{fig:spectraA}), with instances of the synthetic red-shifted component being about twice 
the observed one. This might be related to another discrepancy, namely that the width of the synthetic lines remains sensibly 
smaller than observed, especially for the red-shifted component, even when accounting for Van der Waals and quadratic Stark 
broadening in the calculation (opacity broadening is naturally included in the LTE line synthesis). As the simulations do not include a micro turbulence parameter or reproduce the full range of strong, flare-induced turbulent motions in the chromosphere \citep{2011ApJ...740...70M} it is plausible that flare energy is redistributed into higher intensity, but narrower, line profiles.

\section{Discussion}\label{sec:discussion}

\begin{figure}
  \centering
  \includegraphics[width=9cm]{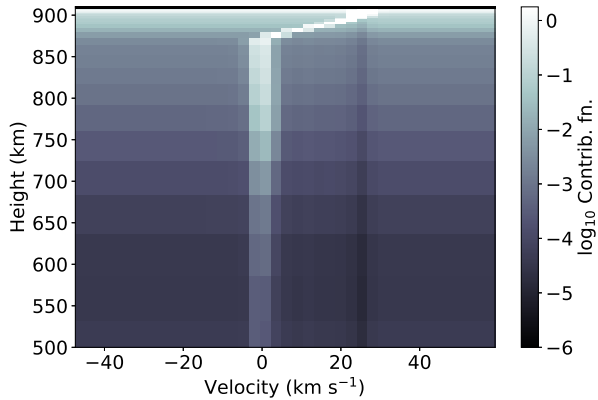}
   \caption{Contribution function (in units of erg cm$^{-2}$s$^{-1}$sr$^{-1}$\AA$^{-1}$cm$^{-1}$)
    of the Fe~{\sc ii} 2814.45\AA~line at 6.5 s within the development
   of the flare. The x-axis is given in units of velocity rather than wavelength. The chromospheric condensation is clearly visible 
   in the upper part of the graph, from $h \sim$ 870--900 km, as the locus of a much enhanced contribution function at longer wavelengths, 
   which gives rise to the red-shifted component.}
  \label{fig:contributionfunction}
\end{figure}

Using the complete description of the physical properties within the simulated flaring atmosphere, we can investigate the origin of the most distinctive observational feature reported above, namely the presence of two separate spectral components for the Fe~{\sc ii} 2814.45\AA~line, each with very different dynamical behaviors. 
%
%
Following the discussion of \citet{2017ApJ...836...12K}, we confirm that these features are due to the concomitant action of accelerated electrons of different energy impinging on the chromosphere. In particular, the highest energy electrons in 
our simulated beam ($ E >$50 keV, increasing to $ E >$80 keV later in the flare evolution)  penetrate the deeper, denser layers of the chromosphere and rapidly heat it to temperatures $\sim 10,000$ K, producing both an enhanced continuum emission and a strongly enhanced emission in the line (``stationary flare layers"). The bulk of the beam's energy, however, 
resides in electrons of lower energy (E $= 15-50$ keV) that are stopped in the higher, more rarefied atmosphere. This causes the development of an explosive chromospheric evaporation and its counterpart condensation, a downward moving high density front ($n_e \sim 2 \times 10^{14}~{\rm cm^{-3}}$), with a thickness of only 30--40 km, and a temperature of 
$T\sim 8000-12000$~K. This layer becomes sufficiently dense as to produce additional emission in both the continuum and the chromospheric spectral lines, resulting in a separate, red-shifted component that traces the downward motion of the condensation towards the stationary chromosphere and its rapid demise. Figure \ref{fig:contributionfunction} illustrates the 
situation at 6.5 s within the flare development: while the stationary component (at v = 0 km s$^{-1}$) is formed within $\sim$
200 km in the mid chromosphere, a very strong contribution appears in the condensation, concentrated in the upper 30--40 km of the atmosphere, at the red-shifted position of v $\sim$ 30 km s$^{-1}$. From the simulations, we find that this moving, dense
 front lasts only a few tens of seconds, as it impacts onto the stationary chromosphere of ever increasing density.

The parameters of the beam of accelerated electrons have an obvious relationship with observable quantities, with the energy cutoff E$_c$, spectral index $\delta$, and total flux F all playing different roles in the creation of the chromospheric signatures,
and of the chromospheric condensation in particular \citep[e.g.][]{2015ApJ...808..177R}. For the case of our flare, as discussed in Sects. \ref{sec:fermi} and \ref{sec:uv_areas}, in the thick target approximation the spectral index of the beam is 
rather well determined, while the low energy cut-off could plausibly vary between 15 and 25 keV; the total flux of 
non-thermal electrons is also determined within about an order of magnitude (cf. Fig. \ref{fig:hxr_fit_parameters}). As an additional
check, we thus ran some further models, keeping constant the spectral index of the beam ($\delta=$ 5) and the
heating duration (20 s), while varying the low energy cutoff (E$_c=$ 15, 18, 20, 25 keV) and the total flux (F = 1, 5 $\times 10^{11} ~{\rm ergs~cm^{-2}~s^{-1}}$; F11, 5F11). The Fe~{\sc ii} 2814.45\AA~ line was then synthesized as described in the previous section.

None of the additional models reproduced the data as closely as the model shown in Fig. \ref{fig:syntheticspectra}, reinforcing the case 
for an electron beam with a relatively low value of $E_c$, and a moderate flux. In particular, for the models with higher cutoff
energy E$_c$ (20-25 keV)
the condensation becomes sensibly weaker, and develops later, with respect to what observed. 
On the other hand, a sensibly higher beam flux (5F11) produces not only a much stronger condensation, but also a much higher continuum 
value than observed (a factor of several), already in the very early instants of the flare. 

The combined diagnostics provided by 
the continuum intensity in the Fe~{\sc ii} line spectral range, and the strength of the condensation, appears thus to be particularly valuable in constraining the model parameters (note that the continuum near the Mg~{\sc ii} lines usually can be determined with much less precision, due to the very extended wings of the line). Still, the second component intensity remains sensibly higher than the stationary one irrespectively of different E$_c$ values, for both the F11 and 5F11 beams. As discussed in the previous section, this is not consistent with observational data; further investigation will be necessary to determine the source of this discrepancy.

\section{Summary and Conclusions}\label{sec:conclusion}

Expanding on our previous work \citep[][Paper I]{Graham:2015aa}, we have studied the details of chromospheric 
dynamics during the impulsive phase of the X1.6 SOL2014-09-10T17:45 event, by employing a comprehensive set of observations and modeling. In particular, the unique set of observations obtained by IRIS
allowed us an unprecedented view of the early impulsive phases of the flare, with large spectral coverage, high cadence (9.4 s) and high spatial resolution ($<$1\arcsec). A novel approach pursued in this work is the use of multiple chromospheric diagnostics available in IRIS spectra, including some weak Fe~{\sc i} and Fe~{\sc ii} lines that are seldom employed. These lines go into clear emission in strong ribbon kernels, but never saturate, contrary to other diagnostics such as the widely used Mg~{\sc ii} lines. Further, the Fe lines remain optically thin during the flare development, hence simplifying their interpretation. 

Our main findings are as follows:

1. Using several, diverse spectral lines  (Figs. \ref{fig:spectraA}, \ref{fig:spectraB}, and \ref{fig:spectraA_16}--\ref{fig:spectraB_52}),
 we confirm that the chromospheric dynamics in the impulsive phase of this flare 
 appears identically in multiple, independent flaring kernels, developing at successive times, separated by as much as several minutes and $\sim$10\arcsec~ (Fig. \ref{fig:ltspec}, bottom panel). This represents a best-case scenario for comparison with (1-D) numerical models of flaring loops.

2. For any given flaring kernel, in the earliest instants of activation all the chromospheric lines show a 
clear double-component structure, with an enhanced spectral line centered at the rest wavelength, and a strongly red-shifted component at $ \Delta \lambda \sim $40 km s$^{-1}$. Their relative intensities and position evolve within a few tens of seconds, with the re-shifted component rapidly 
decelerating until the two components merge in 
an apparently single, very broad and asymmetric line. Such a behavior had been observed sporadically before \citep{2015ApJ...811..139T},
but was never reported so comprehensively during its complete evolution. 

3. The temporal evolution of the redshifted component appears very similar for all spectral lines, and all flaring kernels (Fig. \ref{fig:condensation}), with a timescale of $\sim$ 30 s. This behavior is consistent with the presence of a strong chromospheric condensation, as first modeled by \citet{1985ApJ...289..414F, Fisher:1986aa}. 

4. Adopting the standard thick-target approach, we use co-temporary HXR observations by the Fermi satellite 
 as well as relevant IRIS diagnostics to derive the parameters 
 of a beam of accelerated electrons impinging on the chromosphere. While the beam is not particularly hard ($\delta =5$), we find both a low energy
 cut-off value (E$_c \sim 15 - 20$ keV), and a fairly high energy flux (F =10$^{11}$ erg s$^{-1}$ cm$^{-2}$), which combine
 into a strong heating of the chromosphere. Using these parameters as input to the
 RADYN flare code \citep{1997ApJ...481..500C,2015ApJ...809..104A}, we find that the low chromosphere is rapidly heated
 by the highest energy electrons, while the bulk of the electron beam's energy is dissipated at higher layers. The latter causes the rapid development of an explosive chromospheric evaporation and its counterpart condensation, with maximum velocity $\sim$ 50 km s$^{-1}$.
 The condensation is sufficiently dense to give rise to additional continuum emission, as well as to highly red-shifted components of the analyzed chromospheric lines. The downward motion of the modeled condensation lasts only a few tens of seconds.
 
 5. To properly
 compare the results of the simulation with the actual data, we synthesize the Fe~{\sc ii} 2814.45\AA~line profile in different snapshots of the resulting atmosphere, averaging over time in a manner consistent with the actual IRIS exposures \citep[see also][]{2017ApJ...836...12K}. The synthetic Fe~{\sc ii} profiles (Fig. \ref{fig:syntheticspectra}) reproduce many of the observed characteristics, including the presence of two separate spectral components and their initial separation, as well as the continuum enhancement in the Fe~{\sc ii} window.  As discussed in Sect. \ref{sec:discussion},
the redshifted spectral component and the excess continuum are produced in the condensation described above (point 4), while the stationary component is enhanced because of the heating of the deeper layers due to the penetration of the highest energy electrons. We also find that the 
 continuum intensity in the Fe~{\sc ii} spectral window is an important additional constraint on the details of the energy 
 release, as it is rather sensitive to the beam parameters (Section \ref{sec:discussion}).
The main inconsistencies between the model and the data include a sensibly faster temporal evolution, a larger relative intensity, and a reduced width of the red-shifted component in the simulations, with respect to observations. This could
 be further investigated by adopting a different initial atmosphere in the simulation, and  analyzing the possible role 
 of turbulent flows in the chromospheric condensation, that could possibly influence both the width and intensity of the re-shifted component (cf. Sect. \ref{sec:discussion}).
 


6. For some flaring kernels there are indications of a short ``build-up'' phase towards the maximum red-shift of the condensation component, well visible in the Mg~{\sc ii} 2791.6\AA~panel of Fig. \ref{fig:condensation}. This is not predicted by the hydro-dynamical simulations, and could be related to the 
details of a ``precursor'' phase observed in some instances. However, we have not performed a systematic analysis of the possible effects of noise and/or the instrumental PSF in producing this signature. Recently, using a time dependent, non-equilibrium approach to calculate the atomic level populations,  \cite{2019ApJ...885..119K} have shown that the Mg~{\sc ii} k line becomes slightly red-shifted before the impulsive condensation response (their Fig. 2a). This could be an interesting avenue of further investigation.

As a final curiosity, the slight and short-lived blueshift visible in Fig. \ref{fig:ltspec} after the condensation dies down, appears to represent a rebound of the stationary chromosphere once hit by the condensation, and is reproduced in some RHD simulations like those of \citet[][their Fig. 2]{2016ApJ...827..145R}.

We conclude by remarking that the combination of multiple diagnostics, including HXR emission, UV spectra and continuum intensities, and UV imaging, as well as their temporal evolution, allowed us to strongly constrain the heating and hydrodynamical properties of the impulsive phase of the SOL2014-09-10T17:45 flare. 
%
%
The excellent agreement between multiple observed spectral properties and the results of the 1-D RHD simulations strongly suggests that for this flare we are close to spatially and temporally resolving the impulsive phase of {\it elementary flare kernels}, each one occurring on previously undisturbed chromospheric areas.  In particular, our data does not appear to require any multi-thread scenario of the kind often invoked to explain various flare characteristics, including how the chromospheric emission and/or dynamics of spatially resolved flare foot-points can proceed for a sensibly longer time than predicted by impulsive heating models \citep[e.g.][]{2016ApJ...827..145R,2016ApJ...820...14Q}.

In our study, however, we focused exclusively on the earliest moments of the energization of the chromosphere caused by a strong
electron beam (energy flux $\sim 10^{11}$ erg cm$^{-2}$ s$^{-1}$), and the resulting classical explosive evaporation 
scenario as described by \citet{1989ApJ...346.1019F}. For the longer term chromospheric response, we note that 
the curves of integrated emission of the Fe~{\sc ii} line described in Sect. \ref{sec:duration} and Fig. \ref{fig:ribbon} are tantalizingly similar to the AIA 1600\AA~ curves shown by \citet{2012ApJ...752..124Q} for a different flare (albeit with a vastly different timescale, over 10 times shorter), both showing a sharp peak and a much extended ``gradual'' phase, although new work suggests that care should be taken in interpreting the AIA 1600\AA\ channel \citep{2019ApJ...870..114S}. Whether this behavior is due to the normal cooling of the atmosphere after the flaring episode, or to a two-stage heating process as discussed by \citet{2016ApJ...820...14Q}, will be investigated in a future research.


\acknowledgments
The research leading to these results has received funding from the European Community's Seventh Framework Programme (FP7/2007-2013) under grant agreement no. 606862 (F-CHROMA). DG acknowledges support by NASA under contract NNG09FA40C ({\it IRIS}).   AFK acknowledges support from NASA Helio GI Grant NNX17AD62G.  PJS has received funding from FAPESP (grant 2013/24155-3). IRIS is a NASA small explorer mission developed and operated by LMSAL with mission operations executed at NASA Ames Research center and major contributions to downlink communications funded by the Norwegian Space Center (NSC, Norway) through an ESA PRODEX contract. The National Solar Observatory (NSO) is operated by the Association of Universities for Research in Astronomy, Inc. (AURA), under cooperative agreement with the National Science Foundation. The article benefited from discussions at the International Space Science Institute (ISSI) during a meeting ``Magnetic Waves in Solar Flares: Beyond the Standard Flare Model''. Several of the plots use the colourblind friendly tables provided by \cite{colourpjw}.

\appendix
\section{Additional Spectra}
\label{sec:appendixA}

\begin{figure*}
  \centering
  \includegraphics[width=16cm]{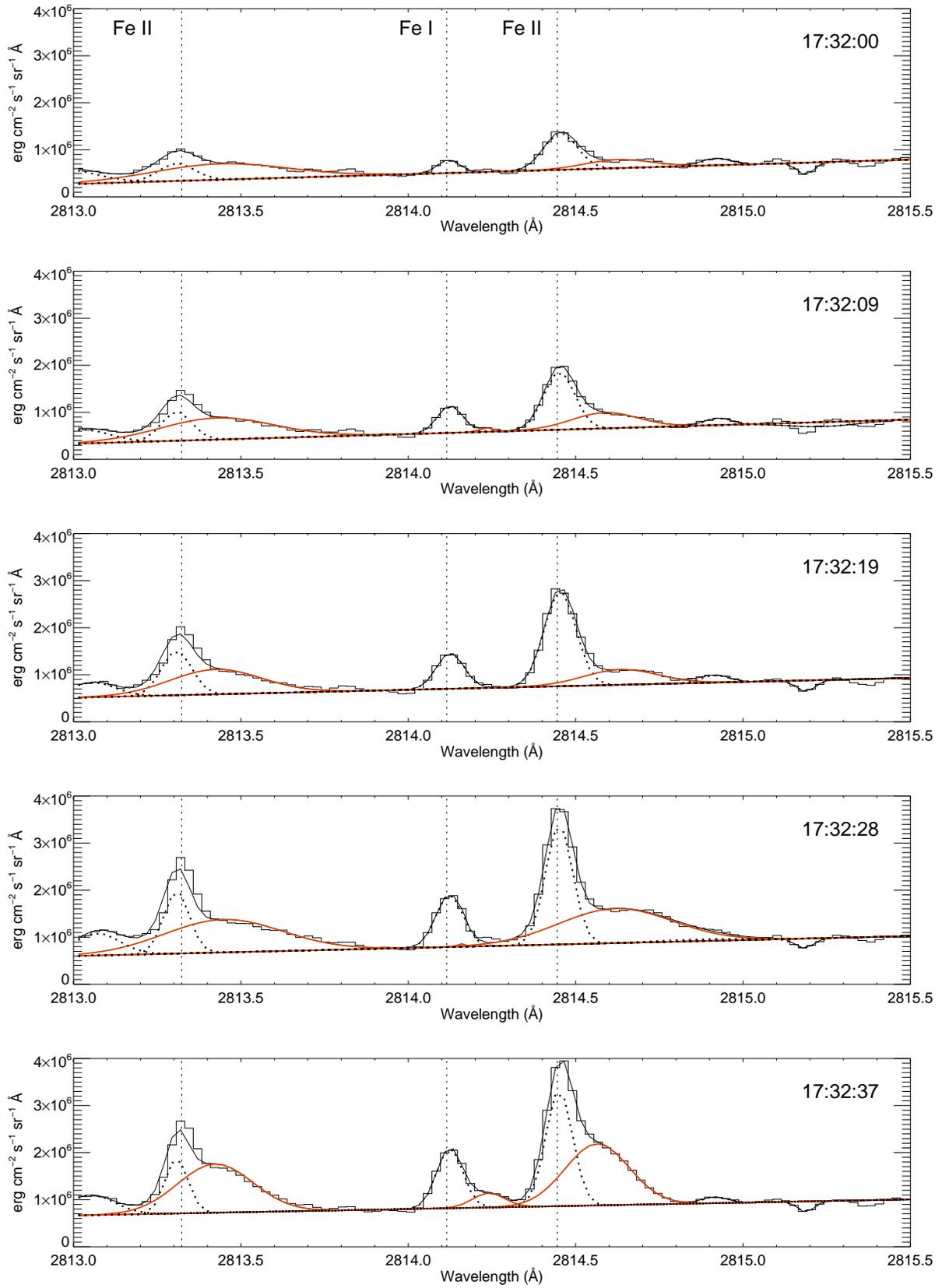}
   \caption{Showing the Fe~{\sc ii} window for multiple chromospheric lines (black, stepped lines) for the y=116.86\arcsec\ slit pixel. Labels as in Fig. \ref{fig:spectraA}}
  \label{fig:spectraA_16}
\end{figure*}

\begin{figure*}
  \centering
  \includegraphics[width=18cm]{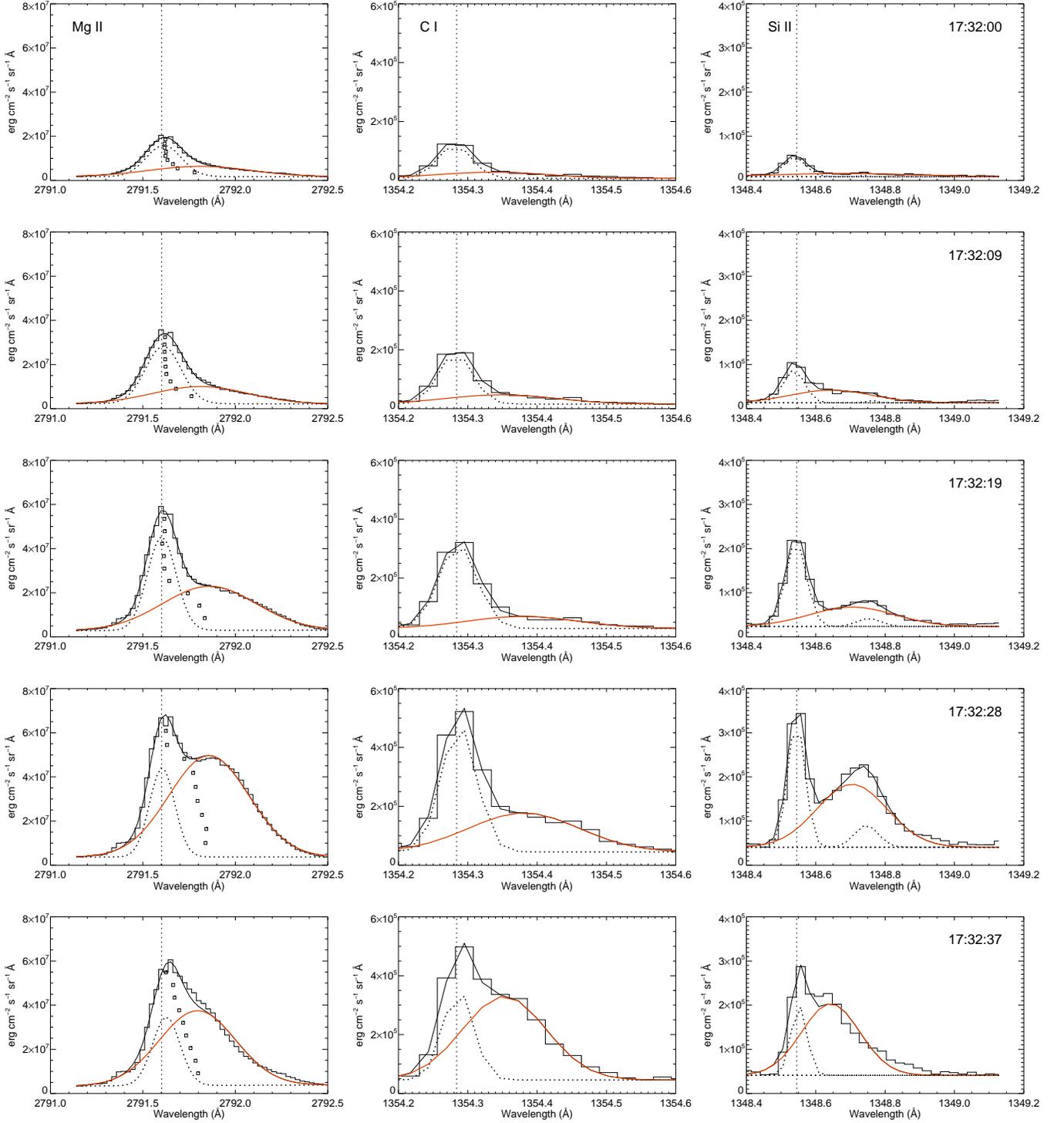}
   \caption{Spectra for the Mg~{\sc ii}, C~{\sc i}, and Si~{\sc ii} lines at y=116.86\arcsec\ slit pixel. Labels as in Fig. \ref{fig:spectraB}}
  \label{fig:spectraB_16}
\end{figure*}

\begin{figure*}
  \centering
  \includegraphics[width=16cm]{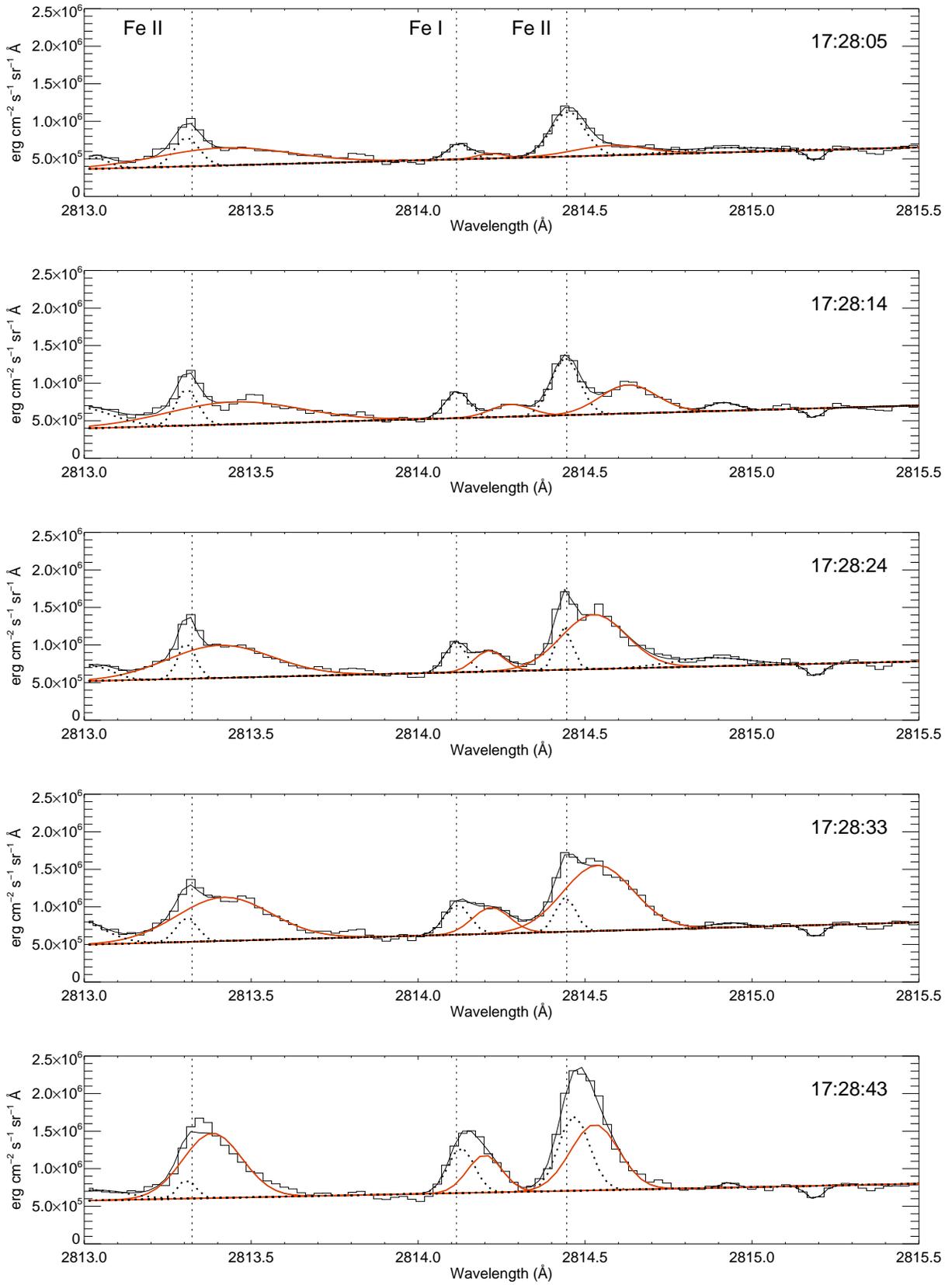}
   \caption{Fe~{\sc ii} window for multiple chromospheric lines (black, stepped lines) for the y=122.86\arcsec\ slit pixel. Labels as in Fig. \ref{fig:spectraA}}
  \label{fig:spectraA_52}
\end{figure*}

\begin{figure*}
  \centering
  \includegraphics[width=18cm]{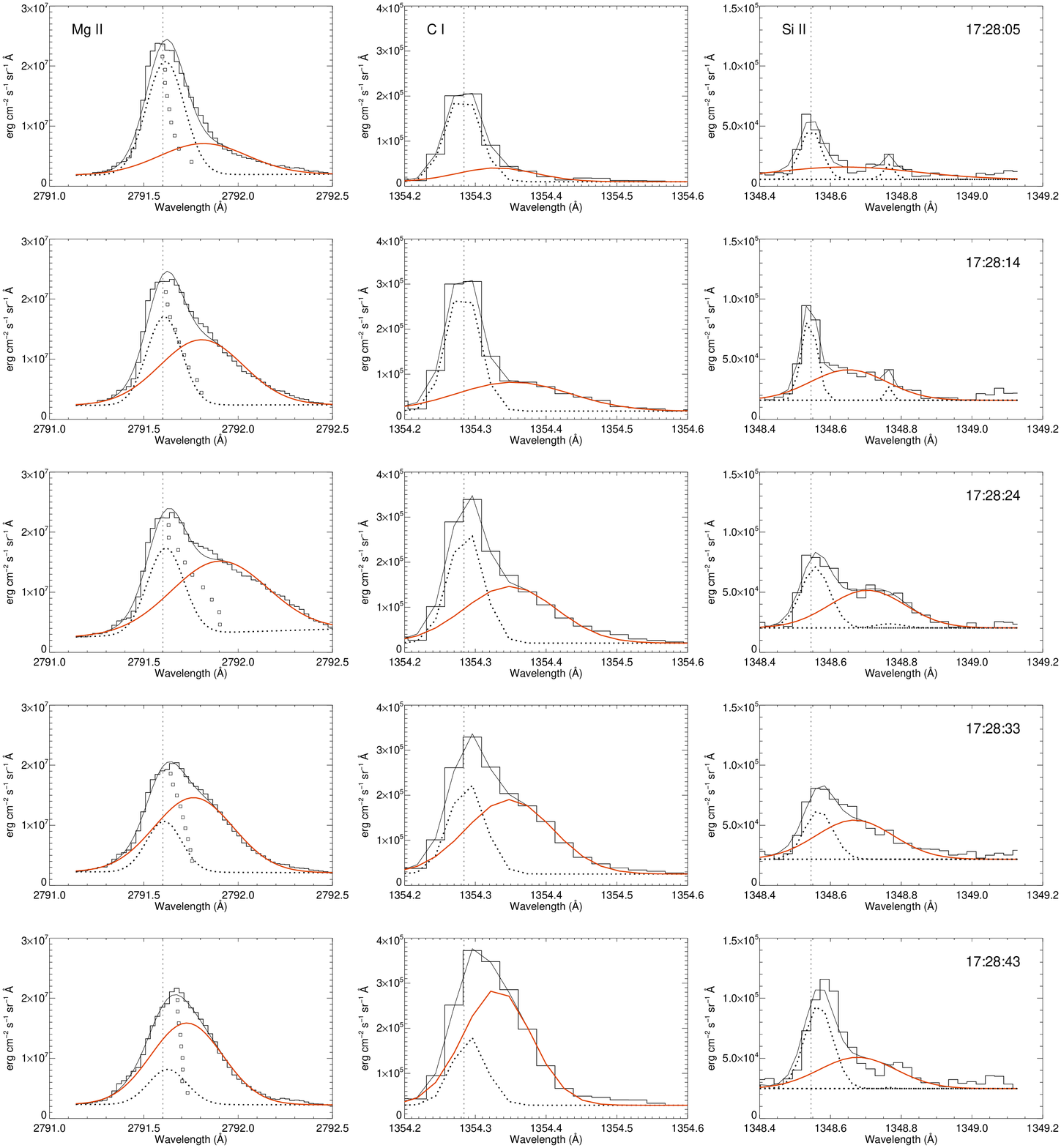}
   \caption{Spectra for the Mg~{\sc ii}, C~{\sc i}, and Si~{\sc ii} lines for the y=122.86\arcsec\ slit pixel. Labels as in Fig. \ref{fig:spectraB}}
  \label{fig:spectraB_52}
\end{figure*}

The spectra shown in Figs. \ref{fig:spectraA} and \ref{fig:spectraB} illustrate a particularly clear example of the presence of a red-shifted component, and its temporal evolution, during the impulsive phase of the flare, but are well representative of many flaring positions along the slit. 
To support the analysis of Section \ref{sec:chrom_spectra} we show in Figs \ref{fig:spectraA_16} -- \ref{fig:spectraB_52}
the corresponding spectra for two additional pixel locations, marked by blue diamonds in the bottom panel of 
Fig. \ref{fig:ltspec}. Each spectral range is shown for five consecutive times starting from the impulsive rise (note the different activation times
in different pixels), and well displays the formation of the red satellite component, and its migration back towards the rest component. The samples are taken from 3.3\arcsec\ and 2.7\arcsec\ above and below the spectra of Figs. \ref{fig:spectraA} and Fig. \ref{fig:spectraB} --- beyond the influence of the point spread function described in Appendix \ref{sec:appendixB} below.

\section{Effects of the IRIS Point Spread Function}
\label{sec:appendixB}

The progression of the ribbon(s) in the  SOL2014-09-10T17:45 flare 
is such that the activation of separate flaring pixels along the slit occurs in a sequential fashion, thus separating in space and time the short-lived kernels, and creating the diagonal intensity streak shown in the lower panel of Figure \ref{fig:ltspec}. 

However, one must be conscious of possible effects of the Point Spread Function (PSF) of the instrument, which could alter the behavior of adjacent pixels, especially in the presence of strong intensity gradients. In particular, an intensity cross-talk along the slit might mimic the activation of a new kernel following a particularly bright kernel, or falsely prolong the duration of a flow in a previously activated pixel. Indeed, close inspection of the space-time plot in Figure \ref{fig:ltspec} shows a general tendency to bright structures being elongated in the slit direction, an effect most apparent for $\sim$1-2\arcsec\ below the brightest pixels at y$=$119\arcsec. 

\cite{2018SoPh..293...20A} report on the stray-light measured along the direction of the slit in observations of the limb, and determine an upper limit to the IRIS PSF of 0.73\arcsec\ (FWHM), i.e. approx. 4-5 IRIS pixels. For the flaring kernels displayed in Fig. \ref{fig:ltspec}, 
the intensity appears to spread further than the nominal NUV spatial resolution of 0.4\arcsec \citep{2014SoPh..289.2733D}, but the smears weaken quickly (in space), compared to the kernel core. For example, for the brightest pixel within the red diamonds' sequence, moving 1\arcsec\ downward along the slit quickly reduces the fitted intensity of the Fe~{\sc ii} line to 38\% of the peak, and to 15\% when moving 2\arcsec\ .

\begin{figure*}
\gridline{\fig{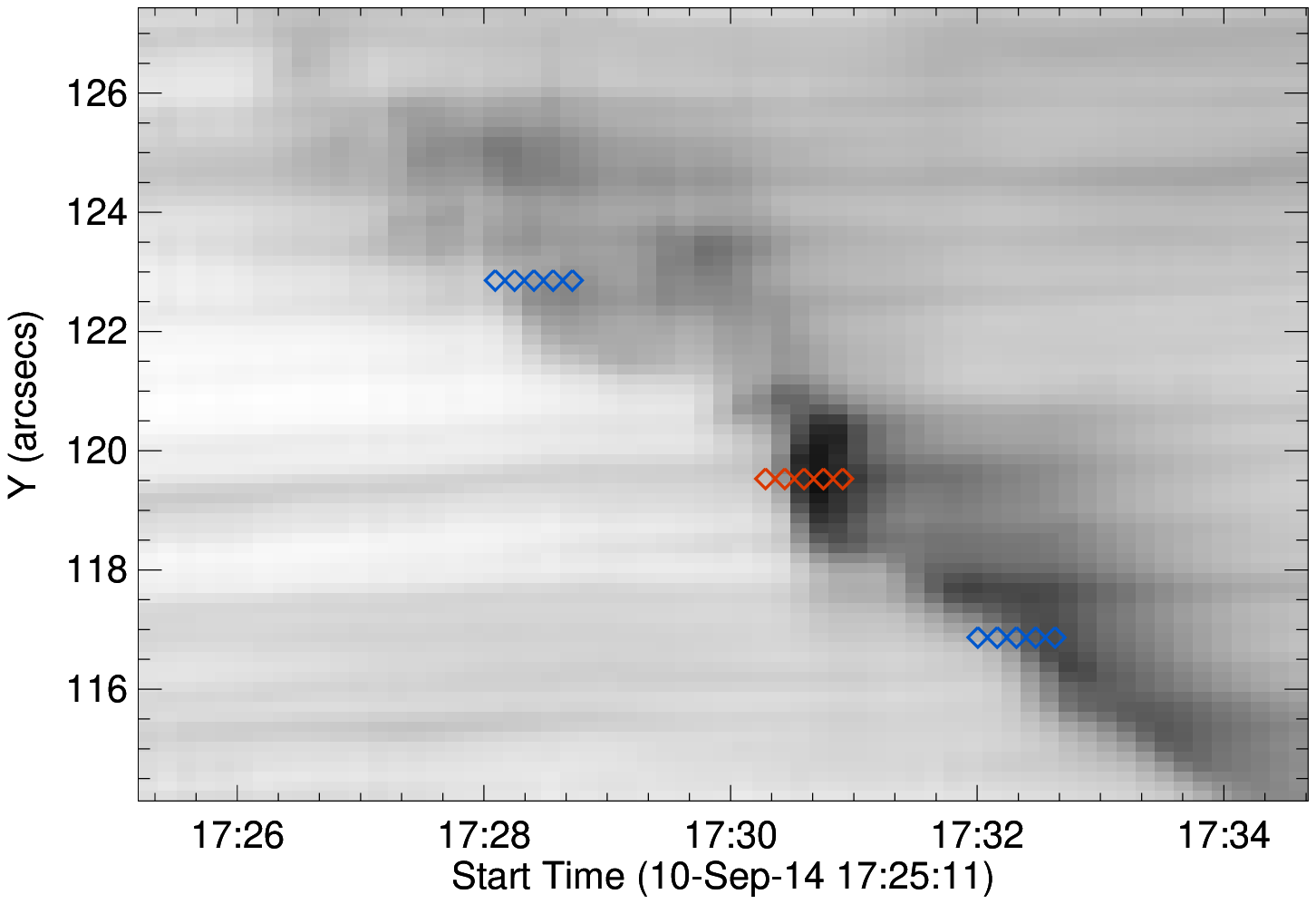}{0.5\textwidth}{(a)}
          \fig{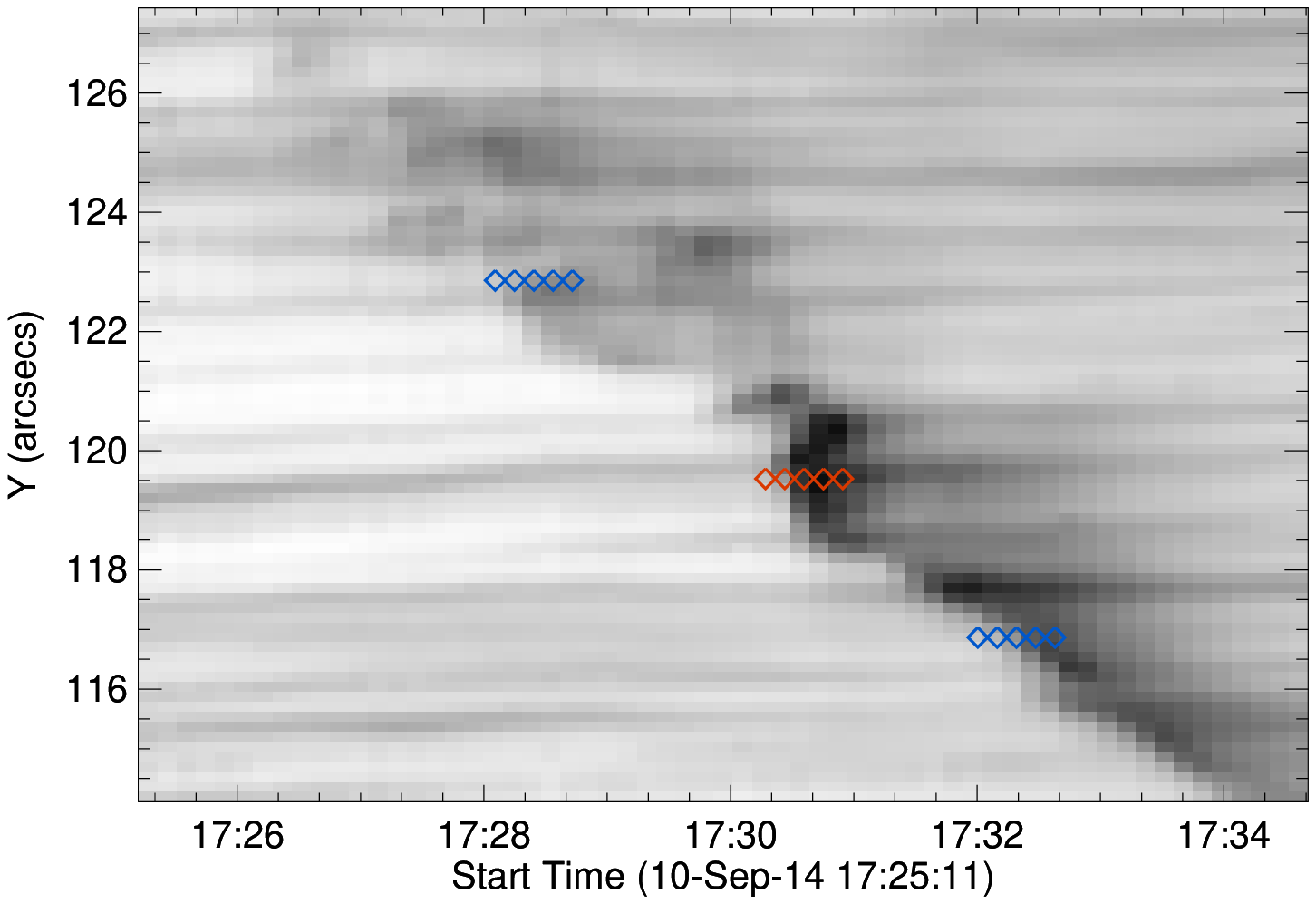}{0.5\textwidth}{(b)}
          }
\caption{Space-time plots of the total Fe spectral window intensity. Panel (a) shows the unfitted observed data and Panel (b) the output from the {\sc iris\_sg\_deconvolve} IDL routine with 15 iterations.
\label{fig:deconv}}
\end{figure*}

Still, we attempted a correction of these effects, using the deconvolution method described by \cite{2018SoPh..293..125C}, who took advantage of a Mercury transit to precisely determine the PSF of IRIS.  Figure \ref{fig:deconv} shows the results. While the deconvolution algorithm clears some of the stray light and sharpens the resulting intensity (Panel (b)), the variations for the flaring pixels are not substantial, especially for what concerns the temporal development. More important, we found that the algorithm introduces additional noise in the spectral profiles, making the complex multi-Gaussian fits that we use quite unstable. Thus, rather than create a new, unquantified source of error, we acknowledge that some amount of stray-light may leak into the rise phase of pixels southward (lower y-values along the slit) of a strong kernel. Given the thresholds used in our fitting routines, and the fact that spectra measured in 
weaker flaring kernels (including those identified with blue diamonds in Fig. \ref{fig:ltspec}, and presented in Figs. \ref{fig:spectraA_16} to \ref{fig:spectraB_52}), are qualitatively and quantitatively very similar to those acquired in the brightest pixel, we believe that this leak has only a minor effect on our results.

\section{Spectral Bisectors}
\label{sec:appendixC}

As described in Sect. \ref{sec:models_results}, and similar to the case discussed in \citet{2017ApJ...836...12K},  the Fe~{\sc ii} 2814.4\AA~line remains optically thin during the flare, thus justifying the use of two separate Gaussians to describe the dynamics and evolution of the different flare components. The fact that a similar analysis appears valid also for the Mg~{\sc ii} 2791.6\AA~line (cf. Fig. \ref{fig:condensation} and Sect. \ref{sec:chrom_time}),
allows us important insights into the use of bisectors, a common method used to estimate the velocity of the condensation 
when using strong, optically thick lines.

Following Ichimoto \& Kurokawa's (\citeyear{1984SoPh...93..105I}) 
explanation of the H$\alpha$ red wing ``excess'' as due to a very broad, red-shifted
component formed in an optically thin condensation, \citet{1990ApJ...348..333C} first introduced the use of 
bisectors, measured in the extreme wings of that line, to derive the amplitude of the motions. 
In their work, \citet{1990ApJ...348..333C} pointed out that the positions of the bisectors should be intensity-independent, as the emission in the
extreme wings would be produced only in the condensation. Yet, for most flares this was not observed,
with the bisectors' position at low intensities usually increasing to ever longer wavelengths. For this reason it became 
customary to estimate the velocity of the condensation using the maximum bisector position \citep{1990ApJ...348..333C},
or that measured at around 20-30\% of the peak line intensity \citep[e.g.][]{1992ApJ...384..341W,Ding:1995aa, Graham:2015aa} , but without providing sound physical reasons for this practice.\footnote{Note that neither
\citet{1984SoPh...93..105I} nor \citet{1990ApJ...348..333C} commented on the strength of the stationary components and its
possible variations.}

The Mg~{\sc ii} panels in Fig. \ref{fig:spectraB} (and Figs \ref{fig:spectraB_16} and \ref{fig:spectraB_52} in  Appendix \ref{sec:appendixA}) offer guidance for the analysis of optically thick, flaring spectral lines. The small squares in the plots identify the bisectors' position measured at 10\% intensity intervals, and illustrate how the bisectors' position can be 
indeed influenced by a complex mix of the parameters of the two spectral components, including
the particular phase of the evolution at which the spectra are acquired. As a general statement, in the earliest stages 
of the condensation the bisectors' position will be observed to increase continuously at lower line intensities, as the 
red-shifted component is quite weaker than the stationary component (top panels in the figures).
On the other hand, when the intensity of the red-shifted component becomes comparable, or even larger, than that of the stationary component, it will dominate the signal in both wings thus 
producing an intensity-independent bisector. This suggests that many of the reported
odd-shapes of strong chromospheric lines in flares are due to the vagaries of the observations - essentially which evolutionary
stage was sampled in which flaring kernel \citep[many flare studies suffer from less than optimal cadence of the
observations. e.g.][]{2015ApJ...806....9K}. 

\begin{figure*}
	\centering
	\includegraphics[width=12cm]{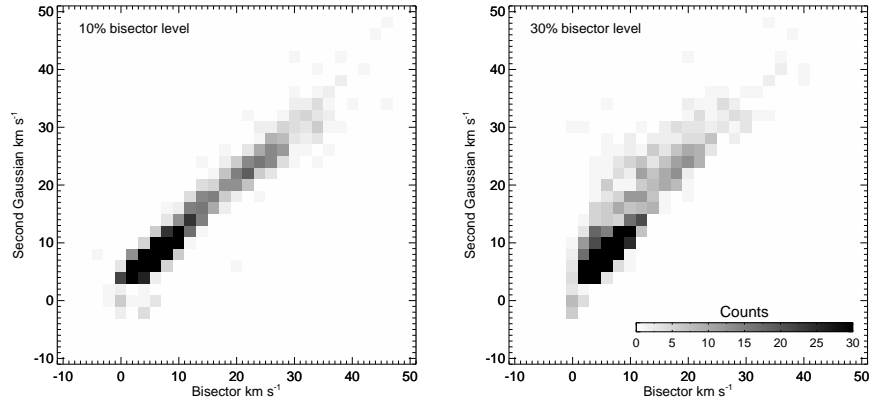}
	\caption{Chromospheric condensation velocities of the Mg~{\sc ii} 2791.6\AA\ line as measured by the bisector, at 10\% and 30\% of the maximum profile intensity, vs the velocity as determined by the 2nd fitted Gaussian component. Correlations 
	between the two quantities are 0.96, and 0.88 respectively (see text). The same selection of fits that appear in Figure \ref{fig:condensation} is used here.}
	\label{fig:bisec_comp}
\end{figure*}

Given the excellent coverage of the impulsive phase of this flare provided by the IRIS data, we are in a unique position 
to test the validity of the bisector method, in comparison with the actual position of the red-shifted component during the flare impulsive phase. Figure 
\ref{fig:bisec_comp} shows a scatterplot of the two quantities for the Mg~{\sc ii} line, comparing the bisector position measured at 10\% (left panel) and 30\% of the overall maximum line intensity (right panel) with the position of the second spectral component derived from the fits. To this end we used the same pixels and spectral fits that enter Fig. \ref{fig:condensation}. The 
gray scale provides the number of sampled spectra that fall within each 2 km s$^{-1}$ bin.

The agreement in Fig. \ref{fig:bisec_comp} is remarkable for the bisector values measured at the 10\% level, with a correlation higher than 0.95. For the 30\% bisector level, the scatter becomes larger, although the correlation remains very high, at 0.88 
(this is driven mostly by the low velocity pixels, identified as the darker bins in the scatterplot; for the pixel/times when the 
measured bisector velocity is $\ge$ 20 km s$^{-1}$, the correlation decreases to 0.75). For higher bisector levels (not 
shown), the correlation with the chromospheric condensation is progressively lost. Thus, we conclude that the bisector values 
can provide a reasonable, {\it lower} limit to the actual condensation velocities, provided they are measured at low-enough 
intensities.

Finally, we note that to date, most studies of chromospheric condensation have been
 conducted using strong, optically thick lines such as  H$\alpha$, Ca~{\sc ii} H\&K, or the Mg~{\sc ii} h\&k \citep[among many,][]{1992ApJ...384..341W,1996A&A...306..625C,1997A&A...328..371F,2015ApJ...806....9K,2015ApJ...811....7L}. 
For such broad lines, most often the two spectral components will appear blended together, as the red-shift of the condensation
does not exceed the natural width of the line, thus giving the appearance of a single, broad, and very asymmetric line (cf. e.g. the last two Mg~{\sc ii} panels in Fig. \ref{fig:spectraB}). This has led to some attempts to interpret the asymmetries as due to a depth-dependent velocity field within the flaring chromosphere \citep[e.g.][]{1996A&A...306..625C}; yet, as shown by \citet{2002A&A...387..678F}, the asymmetries can not be consistently reproduced in models assuming the formation of a single line over a large atmospheric span, even after including an ad-hoc velocity field in the mid-chromosphere. The results of our simulations offer a clear explanation of why this is the case.  
 

\bibliographystyle{apj}

\bibliography{flarebibliogc}

\begin{thebibliography}{86}
\expandafter\ifx\csname natexlab\endcsname\relax\def\natexlab#1{#1}\fi

\bibitem[{{Alissandrakis} {et~al.}(2018){Alissandrakis}, {Vial}, {Koukras},
  {Buchlin}, \& {Chane-Yook}}]{2018SoPh..293...20A}
{Alissandrakis}, C.~E., {Vial}, J.-C., {Koukras}, A., {Buchlin}, E., \&
  {Chane-Yook}, M. 2018, \solphys, 293, 20

\bibitem[{{Allred} {et~al.}(2005){Allred}, {Hawley}, {Abbett}, \&
  {Carlsson}}]{2005ApJ...630..573A}
{Allred}, J.~C., {Hawley}, S.~L., {Abbett}, W.~P., \& {Carlsson}, M. 2005,
  \apj, 630, 573

\bibitem[{{Allred} {et~al.}(2015){Allred}, {Kowalski}, \&
  {Carlsson}}]{2015ApJ...809..104A}
{Allred}, J.~C., {Kowalski}, A.~F., \& {Carlsson}, M. 2015, \apj, 809, 104

\bibitem[{{Allred} {et~al.}(2020){Allred}, {Kowalski}, {Kerr}, \&
  {Alaoui}}]{2020_allred_inprep}
{Allred}, J.~C., {Kowalski}, A.~F., {Kerr}, G.~S., \& {Alaoui}, M. 2020, \apj

\bibitem[{{Bamba} {et~al.}(2017){Bamba}, {Lee}, {Imada}, \&
  {Kusano}}]{2017ApJ...840..116B}
{Bamba}, Y., {Lee}, K.-S., {Imada}, S., \& {Kusano}, K. 2017, \apj, 840, 116

\bibitem[{{Benz}(2017)}]{2017LRSP...14....2B}
{Benz}, A.~O. 2017, Living Reviews in Solar Physics, 14, 2

\bibitem[{{Bissaldi} {et~al.}(2009){Bissaldi}, {von Kienlin}, {Lichti},
  {Steinle}, {Bhat}, {Briggs}, {Fishman}, {Hoover}, {Kippen}, {Krumrey},
  {Gerlach}, {Connaughton}, {Diehl}, {Greiner}, {van der Horst}, {Kouveliotou},
  {McBreen}, {Meegan}, {Paciesas}, {Preece}, \&
  {Wilson-Hodge}}]{2009ExA....24...47B}
{Bissaldi}, E., {von Kienlin}, A., {Lichti}, G., {et~al.} 2009, Experimental
  Astronomy, 24, 47

\bibitem[{{Brosius} \& {Inglis}(2018)}]{2018ApJ...867...85B}
{Brosius}, J.~W., \& {Inglis}, A.~R. 2018, \apj, 867, 85

\bibitem[{{Brown}(1971)}]{1971SoPh...18..489B}
{Brown}, J.~C. 1971, \solphys, 18, 489

\bibitem[{{Canfield} {et~al.}(1990){Canfield}, {Zarro}, {Metcalf}, \&
  {Lemen}}]{1990ApJ...348..333C}
{Canfield}, R.~C., {Zarro}, D.~M., {Metcalf}, T.~R., \& {Lemen}, J.~R. 1990,
  \apj, 348, 333

\bibitem[{{Carlsson} \& {Stein}(1997)}]{1997ApJ...481..500C}
{Carlsson}, M., \& {Stein}, R.~F. 1997, \apj, 481, 500

\bibitem[{{Cauzzi} {et~al.}(1996){Cauzzi}, {Falchi}, {Falciani}, \&
  {Smaldone}}]{1996A&A...306..625C}
{Cauzzi}, G., {Falchi}, A., {Falciani}, R., \& {Smaldone}, L.~A. 1996, \aap,
  306, 625

\bibitem[{{Cauzzi} {et~al.}(1995){Cauzzi}, {Falchi}, {Falciani}, {Smaldone},
  {Schwartz}, \& {Hagyard}}]{1995A&A...299..611C}
{Cauzzi}, G., {Falchi}, A., {Falciani}, R., {et~al.} 1995, \aap, 299, 611

\bibitem[{{Cavallini}(2006)}]{2006SoPh..236..415C}
{Cavallini}, F. 2006, \solphys, 236, 415

\bibitem[{{Cheng} {et~al.}(1981){Cheng}, {Tandberg-Hanssen}, {Bruner}, {Orwig},
  {Frost}, {Kenny}, {Woodgate}, \& {Shine}}]{1981ApJ...248L..39C}
{Cheng}, C.-C., {Tandberg-Hanssen}, E., {Bruner}, E.~C., {et~al.} 1981, \apjl,
  248, L39

\bibitem[{{Cheng} {et~al.}(1988){Cheng}, {Vanderveen}, {Orwig}, \&
  {Tandberg-Hanssen}}]{1988ApJ...330..480C}
{Cheng}, C.-C., {Vanderveen}, K., {Orwig}, L.~E., \& {Tandberg-Hanssen}, E.
  1988, \apj, 330, 480

\bibitem[{{Courrier} {et~al.}(2018){Courrier}, {Kankelborg}, {De Pontieu}, \&
  {W{\"u}lser}}]{2018SoPh..293..125C}
{Courrier}, H., {Kankelborg}, C., {De Pontieu}, B., \& {W{\"u}lser}, J.-P.
  2018, Solar Physics, 293, 125

\bibitem[{{De Pontieu} {et~al.}(2014){De Pontieu}, {Title}, {Lemen}, {Kushner},
  {Akin}, {Allard}, {Berger}, {Boerner}, {Cheung}, {Chou}, {Drake}, {Duncan},
  {Freeland}, {Heyman}, {Hoffman}, {Hurlburt}, {Lindgren}, {Mathur}, {Rehse},
  {Sabolish}, {Seguin}, {Schrijver}, {Tarbell}, {W{\"u}lser}, {Wolfson},
  {Yanari}, {Mudge}, {Nguyen-Phuc}, {Timmons}, {van Bezooijen}, {Weingrod},
  {Brookner}, {Butcher}, {Dougherty}, {Eder}, {Knagenhjelm}, {Larsen},
  {Mansir}, {Phan}, {Boyle}, {Cheimets}, {DeLuca}, {Golub}, {Gates}, {Hertz},
  {McKillop}, {Park}, {Perry}, {Podgorski}, {Reeves}, {Saar}, {Testa}, {Tian},
  {Weber}, {Dunn}, {Eccles}, {Jaeggli}, {Kankelborg}, {Mashburn}, {Pust},
  {Springer}, {Carvalho}, {Kleint}, {Marmie}, {Mazmanian}, {Pereira}, {Sawyer},
  {Strong}, {Worden}, {Carlsson}, {Hansteen}, {Leenaarts}, {Wiesmann},
  {Aloise}, {Chu}, {Bush}, {Scherrer}, {Brekke}, {Martinez-Sykora}, {Lites},
  {McIntosh}, {Uitenbroek}, {Okamoto}, {Gummin}, {Auker}, {Jerram}, {Pool}, \&
  {Waltham}}]{2014SoPh..289.2733D}
{De Pontieu}, B., {Title}, A.~M., {Lemen}, J.~R., {et~al.} 2014, \solphys, 289,
  2733

\bibitem[{{Dickson} \& {Kontar}(2013)}]{Dickson:2013SoPh..284..405D}
{Dickson}, E.~C.~M., \& {Kontar}, E.~P. 2013, \solphys, 284, 405

\bibitem[{{Ding} {et~al.}(1995){Ding}, {Fang}, \& {Huang}}]{Ding:1995aa}
{Ding}, M.~D., {Fang}, C., \& {Huang}, Y.~R. 1995, \solphys, 158, 81

\bibitem[{{Dud{\'{\i}}k} {et~al.}(2016){Dud{\'{\i}}k}, {Polito}, {Janvier},
  {Mulay}, {Karlick{\'y}}, {Aulanier}, {Del Zanna}, {Dzif{\v c}{\'a}kov{\'a}},
  {Mason}, \& {Schmieder}}]{2016ApJ...823...41D}
{Dud{\'{\i}}k}, J., {Polito}, V., {Janvier}, M., {et~al.} 2016, \apj, 823, 41

\bibitem[{{Emslie} {et~al.}(2012){Emslie}, {Dennis}, {Shih}, {Chamberlin},
  {Mewaldt}, {Moore}, {Share}, {Vourlidas}, \& {Welsch}}]{2012ApJ...759...71E}
{Emslie}, A.~G., {Dennis}, B.~R., {Shih}, A.~Y., {et~al.} 2012, \apj, 759, 71

\bibitem[{{Falchi} {et~al.}(1992){Falchi}, {Falciani}, \&
  {Smaldone}}]{1992A&A...256..255F}
{Falchi}, A., {Falciani}, R., \& {Smaldone}, L.~A. 1992, \aap, 256, 255

\bibitem[{{Falchi} \& {Mauas}(2002)}]{2002A&A...387..678F}
{Falchi}, A., \& {Mauas}, P.~J.~D. 2002, \aap, 387, 678

\bibitem[{{Falchi} {et~al.}(1997){Falchi}, {Qiu}, \&
  {Cauzzi}}]{1997A&A...328..371F}
{Falchi}, A., {Qiu}, J., \& {Cauzzi}, G. 1997, \aap, 328, 371

\bibitem[{{Fisher}(1986)}]{Fisher:1986aa}
{Fisher}, G.~H. 1986, in The lower atmosphere of solar flares, p. 25 - 36, ed.
  D.~F. {Neidig}, 25--36

\bibitem[{{Fisher}(1987)}]{Fisher:1987aa}
{Fisher}, G.~H. 1987, \apj, 317, 502

\bibitem[{{Fisher}(1989)}]{1989ApJ...346.1019F}
---. 1989, \apj, 346, 1019

\bibitem[{{Fisher} {et~al.}(1985){Fisher}, {Canfield}, \&
  {McClymont}}]{1985ApJ...289..414F}
{Fisher}, G.~H., {Canfield}, R.~C., \& {McClymont}, A.~N. 1985, \apj, 289, 414

\bibitem[{{Fletcher} {et~al.}(2011){Fletcher}, {Dennis}, {Hudson}, {Krucker},
  {Phillips}, {Veronig}, {Battaglia}, {Bone}, {Caspi}, {Chen}, {Gallagher},
  {Grigis}, {Ji}, {Liu}, {Milligan}, \& {Temmer}}]{2011SSRv..159...19F}
{Fletcher}, L., {Dennis}, B.~R., {Hudson}, H.~S., {et~al.} 2011, \ssr, 159, 19

\bibitem[{{Graham} \& {Cauzzi}(2015)}]{Graham:2015aa}
{Graham}, D.~R., \& {Cauzzi}, G. 2015, \apjl, 807, L22

\bibitem[{{Heinzel} {et~al.}(2016){Heinzel}, {Ka{\v s}parov{\'a}}, {Varady},
  {Karlick{\'y}}, \& {Moravec}}]{2016IAUS..320..233H}
{Heinzel}, P., {Ka{\v s}parov{\'a}}, J., {Varady}, M., {Karlick{\'y}}, M., \&
  {Moravec}, Z. 2016, in IAU Symposium, Vol. 320, Solar and Stellar Flares and
  their Effects on Planets, ed. A.~G. {Kosovichev}, S.~L. {Hawley}, \&
  P.~{Heinzel}, 233--238

\bibitem[{{Heinzel} \& {Kleint}(2014)}]{2014ApJ...794L..23H}
{Heinzel}, P., \& {Kleint}, L. 2014, \apjl, 794, L23

\bibitem[{{Heinzel} {et~al.}(2017){Heinzel}, {Kleint}, {Ka{\v{s}}parov{\'a}},
  \& {Krucker}}]{2017ApJ...847...48H}
{Heinzel}, P., {Kleint}, L., {Ka{\v{s}}parov{\'a}}, J., \& {Krucker}, S. 2017,
  \apj, 847, 48

\bibitem[{{Holman} {et~al.}(2011){Holman}, {Aschwanden}, {Aurass}, {Battaglia},
  {Grigis}, {Kontar}, {Liu}, {Saint-Hilaire}, \&
  {Zharkova}}]{2011SSRv..159..107H}
{Holman}, G.~D., {Aschwanden}, M.~J., {Aurass}, H., {et~al.} 2011, \ssr, 159,
  107

\bibitem[{{Ichimoto} \& {Kurokawa}(1984)}]{1984SoPh...93..105I}
{Ichimoto}, K., \& {Kurokawa}, H. 1984, \solphys, 93, 105

\bibitem[{{Jeffrey} {et~al.}(2018){Jeffrey}, {Fletcher}, {Labrosse}, \&
  {Sim{\~o}es}}]{2018SciA....4.2794J}
{Jeffrey}, N.~L.~S., {Fletcher}, L., {Labrosse}, N., \& {Sim{\~o}es}, P.~J.~A.
  2018, Science Advances, 4, 2794

\bibitem[{{Kennedy} {et~al.}(2015){Kennedy}, {Milligan}, {Allred},
  {Mathioudakis}, \& {Keenan}}]{2015A&A...578A..72K}
{Kennedy}, M.~B., {Milligan}, R.~O., {Allred}, J.~C., {Mathioudakis}, M., \&
  {Keenan}, F.~P. 2015, \aap, 578, A72

\bibitem[{{Kerr} {et~al.}(2019{\natexlab{a}}){Kerr}, {Allred}, \&
  {Carlsson}}]{2019ApJ...883...57K}
{Kerr}, G.~S., {Allred}, J.~C., \& {Carlsson}, M. 2019{\natexlab{a}}, \apj,
  883, 57

\bibitem[{{Kerr} {et~al.}(2019{\natexlab{b}}){Kerr}, {Carlsson}, \&
  {Allred}}]{2019ApJ...885..119K}
{Kerr}, G.~S., {Carlsson}, M., \& {Allred}, J.~C. 2019{\natexlab{b}}, \apj,
  885, 119

\bibitem[{{Kerr} {et~al.}(2015){Kerr}, {Sim{\~o}es}, {Qiu}, \&
  {Fletcher}}]{2015A&A...582A..50K}
{Kerr}, G.~S., {Sim{\~o}es}, P.~J.~A., {Qiu}, J., \& {Fletcher}, L. 2015, \aap,
  582, A50

\bibitem[{{Kleint} {et~al.}(2015){Kleint}, {Battaglia}, {Reardon}, {Sainz
  Dalda}, {Young}, \& {Krucker}}]{2015ApJ...806....9K}
{Kleint}, L., {Battaglia}, M., {Reardon}, K., {et~al.} 2015, \apj, 806, 9

\bibitem[{{Kleint} {et~al.}(2016){Kleint}, {Heinzel}, {Judge}, \&
  {Krucker}}]{2016ApJ...816...88K}
{Kleint}, L., {Heinzel}, P., {Judge}, P., \& {Krucker}, S. 2016, \apj, 816, 88

\bibitem[{{Kontar} {et~al.}(2006){Kontar}, {MacKinnon}, {Schwartz}, \&
  {Brown}}]{Kontar:2006A&A...446.1157K}
{Kontar}, E.~P., {MacKinnon}, A.~L., {Schwartz}, R.~A., \& {Brown}, J.~C. 2006,
  \aap, 446, 1157

\bibitem[{{Kowalski} {et~al.}(2017{\natexlab{a}}){Kowalski}, {Allred}, {Daw},
  {Cauzzi}, \& {Carlsson}}]{2017ApJ...836...12K}
{Kowalski}, A.~F., {Allred}, J.~C., {Daw}, A., {Cauzzi}, G., \& {Carlsson}, M.
  2017{\natexlab{a}}, \apj, 836, 12

\bibitem[{{Kowalski} {et~al.}(2019){Kowalski}, {Butler}, {Daw}, {Fletcher},
  {Allred}, {De Pontieu}, {Kerr}, \& {Cauzzi}}]{2019ApJ...878..135K}
{Kowalski}, A.~F., {Butler}, E., {Daw}, A.~N., {et~al.} 2019, \apj, 878, 135

\bibitem[{{Kowalski} {et~al.}(2015){Kowalski}, {Cauzzi}, \&
  {Fletcher}}]{2015ApJ...798..107K}
{Kowalski}, A.~F., {Cauzzi}, G., \& {Fletcher}, L. 2015, \apj, 798, 107

\bibitem[{{Kowalski} {et~al.}(2017{\natexlab{b}}){Kowalski}, {Allred},
  {Uitenbroek}, {Tremblay}, {Brown}, {Carlsson}, {Osten}, {Wisniewski}, \&
  {Hawley}}]{2017ApJ...837..125K}
{Kowalski}, A.~F., {Allred}, J.~C., {Uitenbroek}, H., {et~al.}
  2017{\natexlab{b}}, \apj, 837, 125

\bibitem[{{Krucker} {et~al.}(2011){Krucker}, {Hudson}, {Jeffrey}, {Battaglia},
  {Kontar}, {Benz}, {Csillaghy}, \& {Lin}}]{2011ApJ...739...96K}
{Krucker}, S., {Hudson}, H.~S., {Jeffrey}, N.~L.~S., {et~al.} 2011, \apj, 739,
  96

\bibitem[{{Kuridze} {et~al.}(2015){Kuridze}, {Mathioudakis}, {Sim{\~o}es},
  {Rouppe van der Voort}, {Carlsson}, {Jafarzadeh}, {Allred}, {Kowalski},
  {Kennedy}, {Fletcher}, {Graham}, \& {Keenan}}]{2015ApJ...813..125K}
{Kuridze}, D., {Mathioudakis}, M., {Sim{\~o}es}, P.~J.~A., {et~al.} 2015, \apj,
  813, 125

\bibitem[{{Li} {et~al.}(2015{\natexlab{a}}){Li}, {Ning}, \&
  {Zhang}}]{2015ApJ...807...72L}
{Li}, D., {Ning}, Z.~J., \& {Zhang}, Q.~M. 2015{\natexlab{a}}, \apj, 807, 72

\bibitem[{{Li} {et~al.}(2015{\natexlab{b}}){Li}, {Ning}, \&
  {Zhang}}]{2015ApJ...813...59L}
---. 2015{\natexlab{b}}, \apj, 813, 59

\bibitem[{{Li} {et~al.}(2015{\natexlab{c}}){Li}, {Ding}, {Qiu}, \&
  {Cheng}}]{2015ApJ...811....7L}
{Li}, Y., {Ding}, M.~D., {Qiu}, J., \& {Cheng}, J.~X. 2015{\natexlab{c}}, \apj,
  811, 7

\bibitem[{{Libbrecht} {et~al.}(2019){Libbrecht}, {de la Cruz Rodr{\'\i}guez},
  {Danilovic}, {Leenaarts}, \& {Pazira}}]{2019A&A...621A..35L}
{Libbrecht}, T., {de la Cruz Rodr{\'\i}guez}, J., {Danilovic}, S., {Leenaarts},
  J., \& {Pazira}, H. 2019, \aap, 621, A35

\bibitem[{{Meegan} {et~al.}(2009){Meegan}, {Lichti}, {Bhat}, {Bissaldi},
  {Briggs}, {Connaughton}, {Diehl}, {Fishman}, {Greiner}, {Hoover}, {van der
  Horst}, {von Kienlin}, {Kippen}, {Kouveliotou}, {McBreen}, {Paciesas},
  {Preece}, {Steinle}, {Wallace}, {Wilson}, \&
  {Wilson-Hodge}}]{2009ApJ...702..791M}
{Meegan}, C., {Lichti}, G., {Bhat}, P.~N., {et~al.} 2009, \apj, 702, 791

\bibitem[{{Milligan}(2011)}]{2011ApJ...740...70M}
{Milligan}, R.~O. 2011, \apj, 740, 70

\bibitem[{{Milligan} {et~al.}(2014){Milligan}, {Kerr}, {Dennis}, {Hudson},
  {Fletcher}, {Allred}, {Chamberlin}, {Ireland}, {Mathioudakis}, \&
  {Keenan}}]{2014ApJ...793...70M}
{Milligan}, R.~O., {Kerr}, G.~S., {Dennis}, B.~R., {et~al.} 2014, \apj, 793, 70

\bibitem[{{Namekata} {et~al.}(2017){Namekata}, {Sakaue}, {Watanabe}, {Asai},
  {Maehara}, {Notsu}, {Notsu}, {Honda}, {Ishii}, {Ikuta}, {Nogami}, \&
  {Shibata}}]{2017ApJ...851...91N}
{Namekata}, K., {Sakaue}, T., {Watanabe}, K., {et~al.} 2017, \apj, 851, 91

\bibitem[{{Ning}(2017)}]{2017SoPh..292...11N}
{Ning}, Z. 2017, \solphys, 292, 11

\bibitem[{{Polito} {et~al.}(2019){Polito}, {Testa}, \& {De
  Pontieu}}]{2019ApJ...879L..17P}
{Polito}, V., {Testa}, P., \& {De Pontieu}, B. 2019, \apjl, 879, L17

\bibitem[{{Qiu} {et~al.}(2010){Qiu}, {Liu}, {Hill}, \&
  {Kazachenko}}]{2010ApJ...725..319Q}
{Qiu}, J., {Liu}, W., {Hill}, N., \& {Kazachenko}, M. 2010, \apj, 725, 319

\bibitem[{{Qiu} {et~al.}(2012){Qiu}, {Liu}, \&
  {Longcope}}]{2012ApJ...752..124Q}
{Qiu}, J., {Liu}, W.-J., \& {Longcope}, D.~W. 2012, \apj, 752, 124

\bibitem[{{Qiu} \& {Longcope}(2016)}]{2016ApJ...820...14Q}
{Qiu}, J., \& {Longcope}, D.~W. 2016, \apj, 820, 14

\bibitem[{{Reep} {et~al.}(2015){Reep}, {Bradshaw}, \&
  {Alexander}}]{2015ApJ...808..177R}
{Reep}, J.~W., {Bradshaw}, S.~J., \& {Alexander}, D. 2015, \apj, 808, 177

\bibitem[{{Reep} {et~al.}(2016){Reep}, {Warren}, {Crump}, \&
  {Sim{\~o}es}}]{2016ApJ...827..145R}
{Reep}, J.~W., {Warren}, H.~P., {Crump}, N.~A., \& {Sim{\~o}es}, P. J.~A. 2016,
  \apj, 827, 145

\bibitem[{{Rubio da Costa} {et~al.}(2016){Rubio da Costa}, {Kleint},
  {Petrosian}, {Liu}, \& {Allred}}]{2016ApJ...827...38R}
{Rubio da Costa}, F., {Kleint}, L., {Petrosian}, V., {Liu}, W., \& {Allred},
  J.~C. 2016, \apj, 827, 38

\bibitem[{{Rubio da Costa} {et~al.}(2015{\natexlab{a}}){Rubio da Costa},
  {Kleint}, {Petrosian}, {Sainz Dalda}, \& {Liu}}]{2015ApJ...804...56R}
{Rubio da Costa}, F., {Kleint}, L., {Petrosian}, V., {Sainz Dalda}, A., \&
  {Liu}, W. 2015{\natexlab{a}}, \apj, 804, 56

\bibitem[{{Rubio da Costa} {et~al.}(2015{\natexlab{b}}){Rubio da Costa}, {Liu},
  {Petrosian}, \& {Carlsson}}]{2015ApJ...813..133R}
{Rubio da Costa}, F., {Liu}, W., {Petrosian}, V., \& {Carlsson}, M.
  2015{\natexlab{b}}, \apj, 813, 133

\bibitem[{{Sandlin} {et~al.}(1986){Sandlin}, {Bartoe}, {Brueckner}, {Tousey},
  \& {Vanhoosier}}]{1986ApJS...61..801S}
{Sandlin}, G.~D., {Bartoe}, J.-D.~F., {Brueckner}, G.~E., {Tousey}, R., \&
  {Vanhoosier}, M.~E. 1986, \apjs, 61, 801

\bibitem[{{Santangelo} {et~al.}(1973){Santangelo}, {Horstman}, \&
  {Horstman-Moretti}}]{Santangelo:1973SoPh...29..143S}
{Santangelo}, N., {Horstman}, H., \& {Horstman-Moretti}, E. 1973, \solphys, 29,
  143

\bibitem[{{Scharmer} {et~al.}(2008){Scharmer}, {Narayan}, {Hillberg}, {de la
  Cruz Rodriguez}, {L{\"o}fdahl}, {Kiselman}, {S{\"u}tterlin}, {van Noort}, \&
  {Lagg}}]{2008ApJ...689L..69S}
{Scharmer}, G.~B., {Narayan}, G., {Hillberg}, T., {et~al.} 2008, \apjl, 689,
  L69

\bibitem[{{Schwartz} {et~al.}(2002){Schwartz}, {Csillaghy}, {Tolbert},
  {Hurford}, {McTiernan}, \& {Zarro}}]{2002SoPh..210..165S}
{Schwartz}, R.~A., {Csillaghy}, A., {Tolbert}, A.~K., {et~al.} 2002, \solphys,
  210, 165

\bibitem[{{Shibata} \& {Magara}(2011)}]{2011LRSP....8....6S}
{Shibata}, K., \& {Magara}, T. 2011, Living Reviews in Solar Physics, 8, 6

\bibitem[{{Sim{\~o}es} {et~al.}(2015){Sim{\~o}es}, {Hudson}, \&
  {Fletcher}}]{2015SoPh..290.3625S}
{Sim{\~o}es}, P.~J.~A., {Hudson}, H.~S., \& {Fletcher}, L. 2015, \solphys, 290,
  3625

\bibitem[{{Sim{\~o}es} \& {Kontar}(2013)}]{Simoes:2013A&A...551A.135S}
{Sim{\~o}es}, P.~J.~A., \& {Kontar}, E.~P. 2013, \aap, 551, A135

\bibitem[{{Sim{\~o}es} {et~al.}(2019){Sim{\~o}es}, {Reid}, {Milligan}, \&
  {Fletcher}}]{2019ApJ...870..114S}
{Sim{\~o}es}, P. J.~A., {Reid}, H. A.~S., {Milligan}, R.~O., \& {Fletcher}, L.
  2019, \apj, 870, 114

\bibitem[{{Tian} \& {Chen}(2018)}]{2018ApJ...856...34T}
{Tian}, H., \& {Chen}, N.~H. 2018, \apj, 856, 34

\bibitem[{{Tian} {et~al.}(2015){Tian}, {Young}, {Reeves}, {Chen}, {Liu}, \&
  {McKillop}}]{2015ApJ...811..139T}
{Tian}, H., {Young}, P.~R., {Reeves}, K.~K., {et~al.} 2015, \apj, 811, 139

\bibitem[{{Tremblay} \& {Bergeron}(2009)}]{2009ApJ...696.1755T}
{Tremblay}, P.~E., \& {Bergeron}, P. 2009, \apj, 696, 1755

\bibitem[{{Uitenbroek}(2001)}]{2001ApJ...557..389U}
{Uitenbroek}, H. 2001, \apj, 557, 389

\bibitem[{{Warren}(2006)}]{2006ApJ...637..522W}
{Warren}, H.~P. 2006, \apj, 637, 522

\bibitem[{{White} {et~al.}(2005){White}, {Thomas}, \&
  {Schwartz}}]{2005SoPh..227..231W}
{White}, S.~M., {Thomas}, R.~J., \& {Schwartz}, R.~A. 2005, \solphys, 227, 231

\bibitem[{Wright(2017)}]{colourpjw}
Wright, P.~J. 2017, {ColourBlind: A Collection of IDL Colour-blind-friendly
  Colour Tables}

\bibitem[{{Wulser} {et~al.}(1992){Wulser}, {Canfield}, \&
  {Zarro}}]{1992ApJ...384..341W}
{Wulser}, J.-P., {Canfield}, R.~C., \& {Zarro}, D.~M. 1992, \apj, 384, 341

\bibitem[{{Zhou} {et~al.}(2016){Zhou}, {Zhang}, \&
  {Wang}}]{2016ApJ...823L..19Z}
{Zhou}, G.~P., {Zhang}, J., \& {Wang}, J.~X. 2016, \apjl, 823, L19

\bibitem[{{Zhu} {et~al.}(2019){Zhu}, {Kowalski}, {Tian}, {Uitenbroek},
  {Carlsson}, \& {Allred}}]{2019ApJ...879...19Z}
{Zhu}, Y., {Kowalski}, A.~F., {Tian}, H., {et~al.} 2019, \apj, 879, 19

\end{thebibliography}

\end{document}